\documentclass[12pt]{article}

\usepackage{graphicx}
\usepackage{color}
\usepackage{bm}
\usepackage{amsmath}

\def\eq#1{(\ref{#1})}

\newcommand{\ket}[1]{|{#1}\rangle}

\def\beq{\begin{equation}}
\def\eeq{\end{equation}}
\def\beqa{\begin{eqnarray}}
\def\eeqa{\end{eqnarray}}
\def\bet{\begin{tabular}}
\def\eet{\end{tabular}}

\newcommand{\ex}[1]{{\rm e}^{#1}} \def\ii{{\rm i}}

\newcommand{\sect}[1]{\setcounter{equation}{0}\section{#1}}

\renewcommand{\a}{\alpha}

\newcommand{\e}{\epsilon}

\def\one{{\hbox{ 1\kern-.8mm l}}}

\begin{document}

\begin{titlepage}

\setcounter{page}{0}

\begin{flushright}
{DFTT 3/2007}\\
{QMUL-PH-06-13}
\end{flushright}

\vspace{0.6cm}

\begin{center}
{\Large \bf The twisted open string partition function\\
and Yukawa couplings} \\

\vskip 0.8cm

{\bf Rodolfo Russo}\\
{\sl 
Centre for Research in String Theory \\ Department of Physics\\
Queen Mary, University of London\\
Mile End Road, London, E1 4NS,
United Kingdom}\\

\vskip .3cm

{\bf Stefano Sciuto}\\
{\sl Dipartimento di Fisica Teorica, Universit\`a di Torino}\\
 and {\sl  INFN, Sezione di Torino}\\
{\sl Via P. Giuria 1, I-10125 Torino, Italy}\\

\vskip 1.2cm

\end{center}

\begin{abstract}
  
  We use the operator formalism to derive the bosonic contribution to
  the twisted open string partition function in toroidal
  compactifications. This amplitude describes, for instance, the
  planar interaction between $g+1$ magnetized or intersecting
  D-branes. We write the result both in the closed and in the open
  string channel in terms of Prym differentials on the appropriate
  Riemann surface. Then we focus on the $g=2$ case for a 2-torus.  By
  factorizing the twisted partition function in the open string
  channel we obtain an explicit expression for the 3-twist field
  correlator, which is the main ingredient in the computation of
  Yukawa couplings in D-brane phenomenological models. This provides
  an alternative method for computing these couplings that does not
  rely on the stress-energy tensor technique.

\end{abstract}

\vfill

\end{titlepage}

\sect{Introduction}
\label{intro}

It is well known that various interesting string vacua are described by free
Conformal Field Theories (CFT) with twisted ({\em i.e.}  periodic up to a
phase) boundary conditions on the world-sheet fields. This happens in the case
of orbifolds for closed strings~\cite{Dixon:1985jw}, and, with the doubling
trick, also for the open strings stretched between
intersecting~\cite{Berkooz:1996km} or magnetized
D-branes~\cite{Abouelsaood:1986gd}. In these cases the 2-dimensional equations
of motions take the usual form of the free wave equation, but the boundary
conditions on the world-sheet fields can change; in particular this happens in
the interactions between closed strings belonging to different twisted sectors
or open strings stretched between different D-branes. The CFT operators
implementing a change in the boundary conditions for the bosonic coordinates
are called twist fields and are rather complicated objects from the
world-sheet point of view. In the following we will use $\sigma_{\e}$ to
indicate a twists field changing the boundary conditions of a complexified
bosonic coordinate by a phase $\ex{2\pi\ii \e}$. In superstring theories, a
similar pattern is present also in the fermionic sector; however in this case
the operators changing the boundary conditions (called spin fields) can be
described in terms of simple free fields thanks to the bosonization
equivalence. In this paper we will exclusively focus on the bosonic twist
fields for which no simple free theory description is known.

The couplings among 3-twist fields are important in various context,
in particular in string compactifications that display realistic
phenomenological features, where they are directly related to the
Yukawa couplings of the low energy effective action. The twist fields
correlators are usually computed by using the techniques introduced
in~\cite{Dixon:1986qv}. In the context of closed string theory on
orbifolds these techniques were applied to the computation of Yukawa
couplings in~\cite{Burwick:1990tu}.
More recently other semi-realistic string vacua have been under
intense study: in this case the standard model fields are realized by
means of open strings stretched between intersecting or magnetized
D-branes. From a CFT point of view, the main ingredient in the
determination of the Yukawa couplings in these models is the
evaluation of a 3-twist coupling. By following the prescription
of~\cite{Dixon:1986qv}, the twist-fields correlators in open string
setups have been studied in detail\footnote{The first analysis of the
twist fields correlators in the open string case was given
in~\cite{Gava:1997jt}} by~\cite{Cremades:2003qj,Cvetic:2003ch}.

Another very important quantity that can be computed in string configurations
with twisted boundary conditions is the partition function. This can be seen
as a contribution to the cosmological constant and thus serves, in orbifold
compactifications, as a check of whether supersymmetry is broken or not in the
twisted sectors~\cite{Kachru:1998pg}. In open string
configurations the partition function captures the D-brane dynamics and string
coordinates with non-trivial monodromies appear when the D-branes support
constant electro-magnetic fields~\cite{Bachas:1992bh} or have constant
velocity~\cite{Bachas:1995kx}. Twisted partition functions are of
interest also in topological string theory and Heterotic strings, see for
example~\cite{Antoniadis:2005sd,Dabholkar:2006bj} for recent applications.

In this paper we use the operator
formalism~\cite{Alvarez-Gaume:1988bg,DiVecchia:1988cy} to compute the twisted
partition function for open bosonic strings in a generic toroidal
compactification. For concreteness, we work in the setup of open strings
stretched between magnetized D-branes, where the parameters $\e_i$ depend on
the magnetic fields on the D-branes. This setup is more general than the one
usually encountered in orbifold compactifications, because the twists $\e_i$
are generically non-rational numbers. In this paper we make two simplifying
assumptions: we require that the D-branes wrap only once the geometrical torus
and impose that the world-volume magnetic fluxes commute, see~\eq{commute}.
While the first assumption provides just a technical simplification, the
presence of non-abelian backgrounds would make the problem rather more
difficult. We hope to relax these hypothesis in a future work.

We will start our computation in the closed string channel, where the
main ingredient is the boundary state describing a magnetized D-brane
(see~\cite{DiVecchia:1999fx} and Refs. therein). We follow the
approach of~\cite{Frau:1997mq} and build a surface with $g+1$ borders
starting from a vertex describing the interaction among $g+1$ closed
strings. The two main novelties in the present computation are the use
of magnetized boundary states and the analysis of the contribution of
winding or Kaluza-Klein modes present in the compact case. Then
by using the results of~\cite{Russo:2003tt,Russo:2003yk} we exploit
the modular properties of string amplitudes to rewrite the same
diagram with magnetized D-branes in the open string channel. All our
expressions are written in terms of geometrical quantities that
commonly appear in the study of Riemann surfaces, such as the period
matrix or the Riemann class. In particular, in our case the
Prym differentials, which are 1-forms with non-trivial monodromies,
play a crucial role. The operator formalism naturally gives for all
these quantities, including the Prym
differentials~\cite{Russo:2003tt}, an explicit expression in terms of
sums or products over the Schottky group. Even if we use as starting
point a setup with magnetized D-branes, we can easily perform a
T-duality at any step of the computation. So, in our approach, the
computations in the intersecting and in the magnetized descriptions
are essentially on the same footing. In principle, for rational twists
$\e_i$, an expression for the twisted partition function could be
derived from the higher genus correlation functions studied
in~\cite{Atick:1987kd}, but, to the best of our knowledge, this has
not been done explicitly.

\begin{figure}
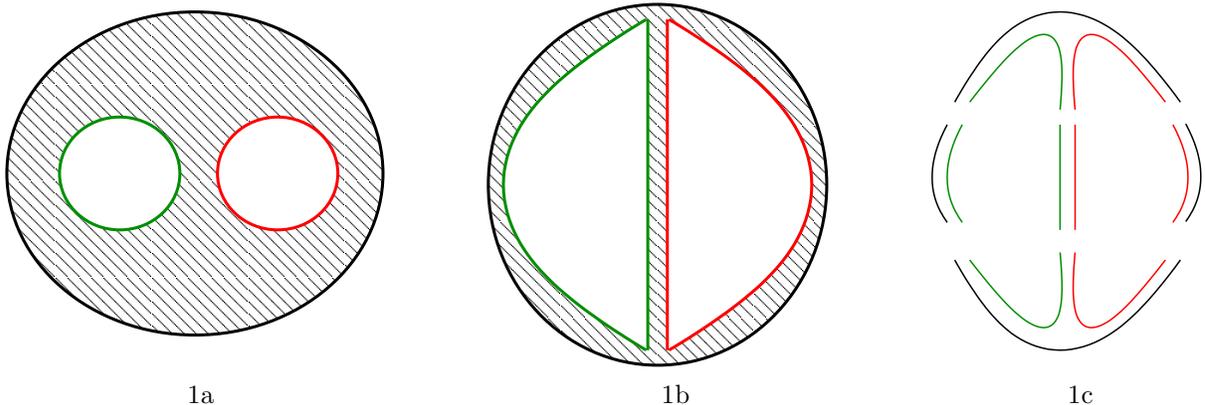

\input fact.pstex_t
\caption{\ref{fact}a represents the twisted 2-loop partition function in the
  open string channel in a generic point of the world-sheet moduli space: on
  the three borders there are different magnetic fields $F_i$; \ref{fact}b is
  the degeneration limit we are interested in, which is obtained by focusing
  on the corner of the world-sheet moduli space defined in~\eq{ftl};
  \ref{fact}c is the factorization of \ref{fact}b into two twisted $3$-string
  vertices and three propagators: this is obtained by focusing only on the
  leading term in the expansion of the previous point \label{fact}.}
\end{figure}

In a second part of the paper we specialize to the $T^2$ case for $g=2$ and
show explicitly that the twisted partition function can be used also to derive
the correlators among three twist fields and thus the Yukawa couplings in
models with magnetized D-branes. The idea is simple: instead of
following~\cite{Dixon:1986qv} and factorizing the 4-point tree-level
amplitude, we derive the couplings among twist fields by factorizing the
planar 2-loop vacuum diagram in the open string channel as depicted in
Fig.~\ref{fact}.
This alternative option was already proposed in~\cite{Friedan:1986ua},
where the formal properties of higher loop amplitudes were
studied. However this program has not been carried out explicitly in
any concrete case and probably the main stumbling point has been so
far the lack of explicit enough formulae for the multiloop twisted
partition function itself. At first sight, the idea of factorizing loop
instead of tree amplitudes may seem an unnecessary
complication. Actually it turns out that for certain purposes it is
technically easier to follow this path. For instance, in the case of
the multiloop partition function, it is easier to control the overall
$\e_i$-dependent normalization, since it is directly related to the
tension of the D-branes. Moreover the computation of the partition
function in our formalism is completely algorithmic and does not require
any space-time intuition.  In fact we will derive the Yukawa couplings
directly in a type IIB setup, thus providing, without the need of any
T-duality, a string counterpart of the field theory computation of
Ref.~\cite{Cremades:2004wa}. However, as we already said, T-dualities
can be easily performed in our approach. So one can see that, in a
type IIA setup, the zero-mode contribution to the partition function
depends on the K\"ahler moduli and, after factorization, reproduces
exactly the world-sheet instanton contribution to the Yukawa couplings
for intersecting brane configurations.

The structure of the paper is the following. 
In Section~2 we summarize the main properties of open and closed
bosonic strings in toroidal compactification and fix our conventions,
see~\cite{Bianchi:1991eu,Giveon:1994fu} for more details.  In
Section~3 we generalize the approach
of~\cite{Russo:2003yk,Magnea:2004ai} and compute the twisted partition
function on a $T^{2d}$ for a planar orientable Riemann surface with
$g+1$ borders and no handles. We first work in the closed string
channel where the amplitudes is obtained by sewing together $g+1$
boundary states on a sphere.  Then we perform a modular transformation
and translate the result in the open string channel. This is done by
using the generalization of the product formulae for the Theta
functions derived in~\cite{Russo:2003tt}. In Section~4 we focus on the
$g=2$ partition function for a 2-torus and study the limit where the
surface degenerates in two disks connected by three open string
propagators. As suggested by Fig.~\ref{fact}, this allows us to derive
the 3-string coupling among twisted states. We first consider the
non-zero mode contribution which is the only one present in the
uncompact space. The contribution to the 3-twist couplings derived in
this way is usually indicated as the quantum (from the world-sheet
point of view) part. Then we study the factorization of the full
twisted partition function on $T^2$. In this way we obtain the full
3-twist couplings, including what is usually called the classical
contribution. Our results for these couplings are in agreement with
previous
works~\cite{Burwick:1990tu,Cremades:2003qj,Cvetic:2003ch,Cremades:2004wa},
thus providing a very non-trivial test for the expression of the
twisted partition function we obtained.

\sect{Open and closed strings on magnetized tori}
\label{conv}

The setup we consider is that of a toroidal compactification $T^{2d}$
of bosonic string theory with $g+1$ space filling D-branes labeled by
the index $i=0,\ldots,g$. The closed strings feel the standard
geometrical metric $G_{MN}$ and the NS-NS $B_{MN}$ field that we will
take to be constant. Our coordinates indicated with the indices
$M,N,\ldots$ are parallel to the lattice defining the torus and have
period $2\pi\sqrt{\a'}$ (this is the so-called integral basis). The
end-points of the open strings are charged under the gauge invariant
combination ${\cal F}_i = B+2\pi\a' F_i$ between the $B$ field and the
$U(1)$ magnetic field $F_i$ living on the D-brane considered. We
restricts ourselves to the case where also the magnetic fields are
constant so that the string world-sheet action is quadratic. Notice
that, in the integral basis, the elements of $F_{i}$ are
quantized
\begin{equation}
\label{qc}
F_{i\;MN}= \frac{1}{2\pi\a'} \frac{p_{i\,MN}}{l_{i\,M}
  l_{i\, N}}~, ~~~~~~i=0,1,..,g,
\end{equation}
where $l_{i\,M}$ is the winding number of the D-brane along the direction
$M$ and $p_{i\,MN}$ is the magnetic flux in the plane $MN$. If an open
string is stretched between the D-branes $k$ and $l$, with gauge
fields $F_k$ and $F_l$, then the boundary conditions on the open
string coordinates $ x^M(z,\bar{z})$ are specified by the reflection
matrix\footnote{We follow the conventions of~\cite{Bertolini:2005qh}}
$R_i$:
\beqa
&\overline\partial x^M(z,\bar{z})\Big|_{\sigma=0}=(R_{k})^M_{~N}\,\partial
x^N(z,\bar{z})\Big|_{\sigma=0}~,~~~~
\overline\partial x^M(z,\bar{z})\Big|_{\sigma=\pi}=(R_{l})^M_{~N}\,\partial
x^N(z,\bar{z})\Big|_{\sigma=\pi}\, &\label{bc-op} \\ \label{R} &
{\rm with}~~~
R_i=\big(G-{\cal F}_i\big)^{-1}\,\big(G+{\cal F}_i\big)~,
\eeqa
where, as usual, $z={\rm e}^{\tau+\ii\sigma}$. Then the mode expansion
of $ x^M(z,\bar{z})$ can be written in terms of a single meromorphic
function $X^M(z)$ whose monodromy properties are encoded by the matrix
$R_{lk}$:
\begin{eqnarray}
& x^M(z,\bar{z}) = q^M + \frac 12 \left[X^M(z) + (R_k)^M_{~N}
  X^N(\bar{z})\right] 
& \nonumber \\ &
X^M({\rm e}^{2\pi{\rm i}}z) =
(R_{lk})^{M}_{~N}\,X^N(z)
~,~~\mbox{with}~~~R_{lk} \equiv R_l^{-1} R_k~.&
\label{monodr}
\end{eqnarray}
When the gauge fields $F_k$ and $F_l$ are non-zero but equal, then we see that
$X^M(z)$ is periodic and we are in the so-called non-commutative setup. In
this case the zero-modes $q^M$ develop non-trivial commutation
relations~\cite{Chu:1998qz} proportional to $(R_l^{MN}-R_l^{NM})/4$ and in
general all open strings modes are contracted with the open string metric:
$G_{\rm open}^{MN}\equiv G^{MN}/2+ (R_l^{MN} + R_l^{NM})/4$. The multiloop
partition function and amplitudes for bosonic strings in this background were
computed in~\cite{Chu:2000wp}.

When $F_k\not= F_l$, the meromorphic function $X^M(z)$ has non-trivial
monodromy properties encoded by $R_{lk}$. Since all $R_i$'s are
$2d\times 2d$ dimensional real matrices satisfying 
\beq\label{orth-pro}
{}^{\rm t}\! R_i\, G \, R_i = G~, ~~~~~~i=0,1,..,g,
\eeq
the eigenvalues of the monodromy matrix $R_{lk}$ are just pairs of complex
conjugate numbers of norm~1.  Thus these eigenvalues can be represented by
phases and we can extract unambiguously $d$ independent parameters that will
be indicated\footnote{The sign ambiguity in the exponent can be fixed by
  requiring for instance $\e_{01}^a \geq 0$.} by $\ex{2\pi\ii
  \e^a_{lk}},~~a=1,2,...,d $.  The mode expansion of $X(z)$ is more easily
written in the (complex) basis where the monodromy matrix is diagonal; we
indicate the meromorphic fields in this basis by ${\cal Z}^a(z)$. The
periodicity properties imply that for the string stretched between the branes
$k$ and $l$ the modes are shifted by $\pm\e^a_{lk}$:
\beq
\partial\mathcal{Z}^{\,a}(z) = -{\rm i}\,\sqrt{2\alpha'}\Bigg( 
\sum_{n=1}^\infty\overline{\a}^a_{n-\e^a_{lk}}\,z^{-n+\e^a_{lk}-1} +
\sum_{n=0}^\infty{\a}^{\dagger\,a}_{n+\e^a_{lk}} 
    \,z^{n+\e^a_{lk}-1} \Bigg)~,~~~a=1,2,...,d, 
\eeq
with commutation relations
\begin{equation}
\big[\overline{\a}^a_{n-\e^a_{lk}}\,,\,
\overline{\a}^{\dagger\,b}_{m-\e^b_{lk}}\big]=(n-\e^a_{lk})\,\delta^{ab}\,
\delta_{n,m}
~~,~~~~ 
\big[{\a}^a_{n+\e^a_{lk}}\,,\,{\a}^{\dagger\,b}_{m+\e^b_{lk}}\big]
=(n+\e^a_{lk})\,\delta^{ab}\,\delta_{n,m}~.
\label{commop}
\end{equation}
The vertex operators describing the physical states of these twisted
open strings contain a product of twist fields $\prod_{a=1}^d
\sigma_{\e^{a}_{lk}}$, each one twisting the string coordinate defined
by the corresponding $R_{lk}$ eigenvector. This implies that the open
string spectrum depends on $G$, $B$ and the $F_i$'s: for instance the
first state (which is the usual open string tachyon in absence of
fluxes) has a mass square given by
\beq\label{twi-tac}
\a' M^2_{lk} = -1 + \sum_{a=1}^d \e_{lk}^a (1-\e_{lk}^a)/2~.
\eeq
The $\e$-dependent shift with the respect to the usual value is expected,
since it is nothing else than the contribution to $L_0$ coming from the
conformal weight of each twist field $\sigma_{\e^{a}_{lk}}$.  The zero-mode
structure in this sector is somewhat subtle: the commutation relations are
$[q^M,q^N] = (F^{-1})^{MN}$ and this yields, in toroidal compactification, a
finite dimensional degeneracy for each string state~\cite{Abouelsaood:1986gd}.
In fact the zero-mode part of the string vertex operators should encode the
same information of the field theoretic wave-functions describing a charged
particle on a magnetized torus, and in field theory one finds, for a fixed
energy, finitely many inequivalent wave-functions~\cite{Cremades:2004wa}. In
our approach we will bypass these subtleties and will not need to deal with
the zero-mode structure of the open string vertices.

Going to the closed string sector, we will use the following conventions:
\beq
x^M_{\rm cl}(z,\bar{z}) = \frac{X_{\rm cl}^M(z) + \tilde{X}^M_{\rm
    cl}(\bar{z})}{2} ~,~~{\rm where}~~~
X_{\rm cl}^M(z) = \mathrm{x}^M -\ii \sqrt{2\a'} \a_0^M
\ln z+\ii \sqrt{2\a'}
\sum_{m\not=0}\frac{\a_m^M z^{-m}}{m}\;,
\eeq
with the commutation relations 
\beq
[\a_m^M,\a_{-n}^N]=n\delta_{n,m} G^{MN}~
~,~~~~~{\rm and}~~~~~~~
[\mathrm{x}^M,\a_0^N] = \ii \sqrt{2\a'} G^{MN}~,
\eeq
and similarly in the right-moving sector. When convenient, we will
 also use the normalized oscillators $a_n$ or $\tilde a_n$, with
 $\a_n=\sqrt{n} a_n$,~ $\a_{-n}=\sqrt{n} a_n^\dagger$ for $n>0$ 
and $\a_0=a_0$.

From the closed string point of view, the D-branes are described by boundary
states $|B_i\rangle$, that enforce an identification between the left and the
right moving modes~\eq{bc-op}. In order to write this identification in the
closed string channel, it is convenient to map the upper-half complex plane
$z$ into the circle of unit radius $w= -(z-\ii)/(z+\ii)$. Thus we get
\beq\label{bc-cl}
{\rm Eq.~\eq{bc-op}} ~~~\Rightarrow~~~ \overline\partial
x^M(w,1/\bar{w})\Big|_{\tau=0}=-(R_{i})^M_{~N}\,\partial
x^N(w,1/\bar{w})\Big|_{\tau=0} ~.
\eeq
At the level of closed string modes this condition reads 
\beq\label{Rcl2}
\left[(G+{\cal F}_i)_{MN} \;\a_n^N + (G-{\cal F}_i)_{MN}
  \;\tilde{\a}_{-n}^N \right] |B_i\rangle=0,~~~~\forall n \in Z~.
\eeq
Notice that the closed strings modes are integer as usual, and the
magnetic fields $F_i$ enter only in the gluing conditions. Thus,
contrary to what happens for the twisted open strings, the closed
string spectrum is unaffected by the magnetic fields. In particular,
the allowed winding and Kaluza-Klein modes are encoded in the Narain
lattice which depends only on $G$ and $B$.  In our conventions the
eigenvalues of the operators $\a_0$ and $\tilde{\a}_0$ are
\beq\label{nar-lat}
(\a_0)^M \!=\! \frac{G^{MN}}{\sqrt{2}}\left[{n_N} + (G_{NN'}- B_{NN'})
  m^{N'}\right]
  ~,~~
(\tilde{\a}_0)^M =  \frac{G^{MN}}{\sqrt{2}}\left[{n_N} - (G_{NN'}+ B_{NN'})
  m^{N'}\right]. 
\eeq
In terms of the Kaluza-Klein and windings mode, the identification~\eq{Rcl2}
enforced by the boundary state (for $n=0$) is independent of the closed string
background and reads simply
\beq\label{Rcl3}
n_M = -2\pi \a' (F_i)_{MN} m^N~.
\eeq

Let us end our brief comments on the string spectrum on magnetized tori by
noticing that it is easy to generalize this analysis to all other
configurations that are connected by T-duality to the case of magnetized
D-branes.  A generic T-duality transformation is encoded in a $O(d,d,Z)$
matrix 
\beq\label{T}
T = \left(\begin{array}{cc}
a_M^{~N} &b_{MN} \\ c^{MN} &d^{M}_{~N}
\end{array}\right)~,~~~~~{\rm with}~~~ {}^{\rm t}\!c\,a +{}^{\rm
  t}\!a\,c = {}^{\rm t}\!b\,d+ {}^{\rm t}\!d\,b =0
  ~~,~~ {}^{\rm t}\!a\,d+^{\rm t}\!c\,b=1.
\eeq
where $a$, $b$, $c$ and $d$ are $2d\times 2d$ matrices.
The matrix $T$ acts on the combination $G+B$ as follows (see~\cite{Giveon:1994fu}
for a review\footnote{One has to take into account that in our conventions
  $\a_n$ and $\tilde{\a}_n$ are exchanged with respect
  to~\cite{Giveon:1994fu}})
\beq\label{sodd}
(G'+B') = \left[a (G+B) + b\right] \left[c (G+B) + d\right]^{-1}\,.~
\eeq

The same T-duality transformation is realized at the level of oscillators by
multiplying the left and right-moving modes by specific matrices: $\a_n\equiv
T_+ \a_n' $, $\tilde{\a}\equiv T_- \tilde{\a}'_n $, where the primed
oscillators are those obtained after the T-duality and
\beq\label{Tpm}
(T_+)^M_{\;N} = \left\{\left[d + c \;(G+B)\right]^{-1}\right\}^M_{~N}
~~,~~~~~
(T_-)^M_{\;N} = \left\{\left[d - c\;{}^{\rm t}\!(G+B)
\right]^{-1}\right\}^M_{~N}~.
\eeq
Notice that by using~\eq{T} one can prove
that ${}^{\rm t}\!T_\pm\, G T_\pm = G'$.
From~\eq{bc-op} we immediately see that upon T-duality the reflection
matrices change in a very simple way $R_i \to {R}'_i = T^{-1}_- R_i
T_+$.  This implies that the monodromy matrices $R_{lk}$ change by a
similarity transformation and that the twists $\e_{lk}$ are invariant
under T-duality. In particular, we can easily describe within the
formalism of reflection and monodromy matrices also D-branes at
angles. This situation is characterized by symmetric reflections
matrices that square to one.  If all $R_i$'s can be brought in this
form by means of the same T-duality transformation $T$, then it means
that the D-brane setup can be geometrized completely and described in
terms of D-branes at angles.

\sect{Twisted partition function on $T^{2d}$}
\label{pf}

\subsection{Twisted partition function in the closed string channel}
\label{clo-chan}

The vacuum diagram we will compute has $g+1$ borders on different D-branes and
no handles or crosscaps. In order to avoid the subtleties of the case with
multiple wrappings~\cite{Pesando:2005df}, we will take all
$l_{iM}$'s in~\eq{qc} to be equal one. Moreover we will take a second
simplifying hypothesis and focus on the case of commuting fluxes. In formulae,
our setup will satisfy
\beq \label{commute}
l_{iM} = 1~,~~~{\rm and}~~~
[G^{-1} {\cal F}_i,G^{-1} {\cal F}_j] = 0~,~~~\forall i,j=0,\ldots,g~.
\eeq
where $G$ and ${\cal F}_i$ are viewed as space-time matrices. Commutativity
implies that all reflection matrices $R_i$ can be diagonalized simultaneously
and so the eigenvalues $\e^a_{lk}$ of the monodromy matrices  $R_{lk}$
 are simply $\e^a_{lk}=e^a_k-e^a_l,~~a=1,2,...,d $, where $\ex{2\pi\ii
  e^a_i}$ are the eigenvalues of $R_i$.

\begin{figure}
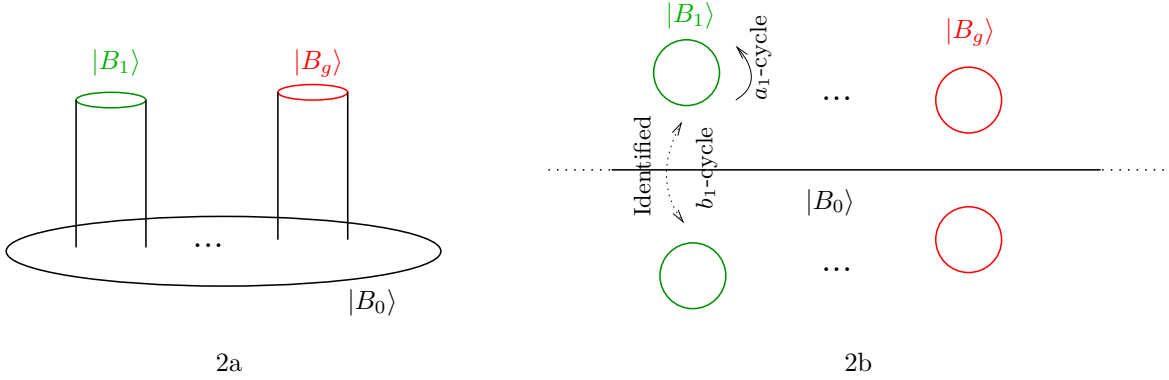

\input clos-par.pstex_t
\caption{\ref{clos-par}a represents the partition function under study
  from a space-time point of view; \ref{clos-par}b is a representation
  from the world-sheet point of view: the world-sheet is the upper half part
  of the complex plane that is outside all disks. \label{clos-par}}
\end{figure}

The computation of this partition function in the closed string channel can be
performed by using the boundary states formalism.  This approach was used
in~\cite{Frau:1997mq} for the case of unmagnetized D-branes and here we will
follow the same steps, the main difference being that in the present case the
reflection matrices $R_i$ contain generic phases and not just $\pm 1$. If we
start from an off-shell closed string vertex in the $z$
coordinates~\eq{close-vert} describing the emission of $g+1$ states and insert
a first boundary state, then we get a disk parametrized as the upper half
complex plane, where the real axis represents the boundary we have just
inserted. On the contrary if we had worked in the $w$ coordinates~\eq{bc-cl},
we would have obtained the disk of unit radius.  Let us focus on the $z$
coordinates and saturate the remaining $g$ off-shell states with other
boundary states.  Each insertion cuts out a small disk in the upper half
complex plane plus an image in the lower part and the circle and its image are
identified. These non overlapping disks are completely specified by the
$2\times 2$ matrices
\beq\label{Scl}
S_\mu^{\rm cl} = \left(\begin{array}{cc}
a_\mu^{\rm cl} &b_\mu^{\rm cl} \\ c_\mu^{\rm cl} &d_\mu^{\rm cl}
\end{array}\right) \equiv \frac{1}{2\sqrt{q_\mu}(\chi_\mu-\bar\chi_\mu)}
\left(\begin{array}{cc}
\chi_\mu - q_\mu\bar{\chi}_\mu& -|\chi_\mu|^2 (1-q_\mu)\\
(1-q_\mu)& q\chi_\mu - \bar{\chi}_\mu
\end{array}\right)~,
\eeq
with $\mu=1,\ldots,g$. The centers of the disks are in $a_\mu/c_\mu$ (and
$-d_\mu/c_\mu$ for the images) and the radii (common to the disks and their
images) are $\sqrt{q_\mu} |\bar{\chi}_\mu-\chi_\mu|/(1-q_\mu)$. Thus we get a
world-sheet parameterization as in Fig.~\ref{clos-par}b, with $g+1$ borders
which are the real axis plus the $g$ circles specified above.
The (real) multipliers $q_\mu$'s and the (complex) fixed points $\chi_\mu$'s of 
Eq.~\eq{Scl} yield $3g$ real parameters; as
usual, three real parameters among the $\chi_\mu$'s can be fixed arbitrarily
thanks to the $SL(2,R)$ invariance, that in this context is simply the freedom
to change all $S_\mu^{\rm cl}$ by a similarity transformation with a $SL(2,R)$
matrix. So we get the correct dimension ($3g-3$) for a the moduli space of a
disk with $g+1$ boundaries. Notice that the parameters $q_\mu$ are directly 
related to the eigenvalues of the $S_\mu^{\rm cl}$'s
(which are $\{\sqrt{q_\mu},1/\sqrt{q_\mu}\}$) and so cannot be fixed by
exploiting the $SL(2,R)$ invariance. The infinite group freely generated by the 
$g$ matrices~\eq{Scl} is called Schottky group (for a disk with $g+1$ boundaries). 
A generic element will be indicated by $T_\alpha^{\rm cl}$ and has the same form of the
generators~\eq{Scl}, so for each $T_\alpha^{\rm cl}$ we can derive a real $q_\a$ and a complex
$\chi_\a$, which are calculable functions of the moduli $q_\mu$, $\chi_\mu$. In
the operator formalism the interaction vertices are written in terms of a
simple representation of the Schottky group~\cite{DiVecchia:1988cy}. Sewing
together vertices and boundary states amounts to multiply together
representations of Schottky elements, so it comes to no surprise that the
final results for the partition function is given in terms of sums and
products over the Schottky group. In the case of interest for us, the sewing
procedure carry also a dependence on the space-time matrices $R_i$ contained
in the boundary states. Here we will recall only the main features of the
computation and the final result, while we leave to the Appendix~\ref{appa}
the discussion of some technical steps.

For the non-zero mode contributions ($\a_n$ and $\tilde\a_n$, with $n\not=0$)
there is a nice pairing throughout the computation between the Schottky
matrices $S_\mu^{\rm cl}$ and the spacetime matrices ${\cal S}_\mu \equiv
R_{0\mu}$: each $(S_\mu^{\rm cl})^{\pm 1}$ is accompanied by the corresponding
${\cal S}_\mu^{\pm 1}$.  In our case all ${\cal S}_\mu$'s can be diagonalized
simultaneously, thus the non-zero mode contribution to vacuum diagram can be
written in terms of the eigenvalues of $T^{\rm cl}_\a$ and those of ${\cal
  S}_\mu$. The first ones are basically the multipliers $q_\a$, while the
second ones are just ${\rm e}^{\pm 2\pi\ii \e^{a}_\mu}$, with
$\e^{a}_\mu\equiv \e^{a}_{0\mu} = e^{a}_\mu -e^{a}_0$, for $\mu=1,2,...,g $.
Notice that from the world-sheet point of view the eigenvalues of ${\cal
  S}_\mu^{\pm 1}$ are simply parameters that twist the periodicity conditions
of the string coordinates along the $b^{\rm cl}$-cycles, that are those
connecting each circle in Fig.~\ref{clos-par} with its own image. This twist
is easily tractable within the operator formalism~\cite{paolo,Russo:2003tt}.
In formulae, the integrand for the twisted partition function can be written
in terms of the $g$-loop untwisted result (all $F_i=0$) times $d$ factors of
$[{\cal R}^a \left(\vec{\e}^a\right)]_g^{\rm cl}$ defined as follows
\beq
[{\cal R}^a \left(\vec{\e}^a\right)]_g^{\rm cl} =  \frac{{\prod_\alpha}' 
\prod_{n = 1}^\infty (1 - q_\alpha^n)^2}{ {\prod_\alpha}' 
\prod_{n = 1}^\infty \left( 1 - \ex{ - 2 \pi \ii \vec{\e}^a \cdot \vec{N}_\a} 
q^n_\alpha \right) \left( 1 - \ex{ 2 \pi \ii \vec{\e}^a \cdot \vec{N}_\a} 
q^n_\alpha \right)}~.
\label{ratioclosed}
\eeq 
In the vector $\vec{\e}^a$ we have collected the $\e^a_{\mu},
~\mu=1,2,...,g$; also $\vec{N}_\a$ is a vector with $g$ integer
entries: the $\mu^{\rm th}$ entry counts how many times the Schottky
generator $S_\mu^{\rm cl}$ enters in the element of the Schottky group
$T_\a^{\rm cl}$, whose multiplier is $q_\a$ (for example $S_\mu^{\rm
cl}$ contributes~$1$, while $(S_\mu^{\rm cl})^{-1}$
contributes~$-1$). The primed product over the Schottky group means
product over primary classes, that is one has to take only the
elements that cannot be written as power of another element of the
group; moreover one has to take only one representative for each
conjugacy class (class of elements that are related by cyclic
permutation of their constituents factors).  In the non-compact case,
the oscillator modes provide the only non-trivial contribution to the
twisted partition function~\cite{paolo}
\beq
Z_g^{\rm cl} (F)|_{\rm uncomp.} = \left( \prod_{i = 0}^g \sqrt{{\rm
    Det}(1-G^{-1} {\cal F}_i)} 
\right)
\int \left[ d Z \right]_g^{\rm cl} \, 
\prod_{a=1}^d [{\cal R}^a \left(\vec{\e}^a\right)]_g^{\rm cl}~,
\label{clch}
\eeq
where $\left[d Z \right]_g^{\rm cl}$ is the integrand of the usual ($F_i = 0$)
partition function in Minkowski space and the superscript $cl$ is just to
recall that we are working in the closed string channel.  Notice that the
overall coefficient is just the usual Born-Infeld Lagrangian rescaled by a
factor of $\sqrt{G}$, since all $F_i$-independent normalizations are included
in $\left[d Z \right]_g^{\rm cl}$. Eq.~\eq{clch} provides a direct
generalization of the $1$-loop result~\cite{Metsaev:1987ju} to the multiloop
case. In fact in this case there is only one conjugacy class and the product
${\prod_\alpha}'$ is absent. So~\eq{ratioclosed} reduces to
\beq
[{\cal R}^a\left({\e}^a\right)]_{g=1}^{\rm cl}  = 
-2\sin(\pi\e^a) \frac{\eta^3(\tau^{\rm
    cl})}{\theta_{11}(\e^a,\tau^{\rm cl})}~, 
\eeq
where $\eta$ and $\theta_{11}$ are the usual Dedekind and odd Theta
function with $q=\ex{2\pi\ii \tau^{\rm cl}}$.

Let us now consider the zero-mode contributions that are present in toroidal
compactification. Each boundary state $\ket{B_i}$ contains a Kronecker delta
selecting a particular combination of winding and Kaluza-Klein modes that
couples to the D-brane, see Eq.~\eq{Rcl3}. The closed string interaction
vertex~\eq{close-vert} contains two independent Kronecker delta's over winding
and Kaluza-Klein modes. Thus the zero-mode contribution arises only from those
modes that satisfy simultaneously all these conditions. By expressing all
$\tilde{\a}_0$ in terms of $\a_0$, one can check that the zero-modes
must satisfy the following condition:
\beq\label{df-cl}
\sum_{\mu=1}^g\left(1-{\cal S}_\mu\right){\a}_0^\mu = 0~.
\eeq
In a long but straightforward computation one can follow the zero-mode
contributions in the computation where the closed string
vertex~\eq{close-vert} is saturated with the boundary states. This
basically amounts to follow what was done in the Appendix~D
of~\cite{DiVecchia:1988cy} and take into account the modifications
induced by the space-time matrices ${\cal S}_\mu$.  The tricky point
is that the pairing between these ${\cal S}_\mu$ and the Schottky
generators $S_\mu^{\rm cl}$ is spoiled in this part of the
computation. This is due to~\eq{D-rep} which implies that the matrices
$D_{nm}$ appearing in the vertex and in the boundary states do not
provide a true representation of the Schottky group in the zero-mode
sector.  In the Appendix~\ref{appa} we provide some details of this
computation, here let us give the final result for the zero-mode
sector written in terms of the eigenvalues $\a_0$ of the left-moving
momentum
\beq\label{zm-cl}
\exp\left\{\frac{1}{2}\sum_{\mu,\nu=1}^g \a_0^\mu G C^{(1)}_{\mu\nu}
\a_0^\nu\right\} = 
\exp\left\{\frac{1}{2}\sum_{\mu,\nu=1}^g \a_0^\mu G \left[{\cal S}_\mu^{-1}
  \int\limits_w^{S^{\rm cl}_\mu(w)} \!\!\bm{\zeta}_\nu(z) dz\right] 
  \a_0^\nu\right\}~.
\eeq
The $\bm{\zeta}_\nu(z)$'s are Prym differential, {\em i.e} closed meromorphic
1-forms, periodic along the $a_\mu^{\rm cl}$ cycles and with fixed monodromies
along the $b_\mu^{\rm cl}$-cycles (see Fig~\ref{clos-par})
\beq\label{perio}
\bm{\zeta}_\nu
\big(S^{\rm cl}_\mu(z)\big) d\big(S^{\rm cl}_\mu(z)\big) =
 {\cal S}_\mu \bm{\zeta}_{\nu}(z) dz ~.
\eeq
The sewing procedure gives an explicit expression for these
1-forms in terms of a sum over the Schottky group:
\beq\label{pe-b-cl}
\bm{\zeta}_\nu(z) = \sum_\a {\cal T}_\a {\cal S}_\nu   
\left[\frac{1}{z - T^{\rm cl}_\a S^{\rm cl}_\nu(z_0)}-\frac{1}{z - T^{\rm cl}_\a(z_0)}
\right]~,
\eeq
where ${\cal T}_\a$ is the space-time counterpart of $T_\a$, with each $S^{\rm
  cl}_\mu$'s replaced by a ${\cal S}_\mu$'s.  These Prym differentials were
already found in~\cite{Russo:2003tt} in the study of the twisted determinants
for the Dirac operator, or in other words the partition function of a
fermionic $b,c$ system of conformal weight $(1,0)$. The same pattern appears
also in the untwisted case with the usual Abelian differentials (which are the
limit of the $\bm{\zeta}_\mu(z)$'s when all ${\cal S}_\mu=1$) and the reason
is obviously that we are dealing with object of conformal weight~$1$ both in
the case of the above mentioned $b,c$ system and in that of the string
coordinates $\partial X$.  Riemann-Roch theorem implies that it is not
possible to find $g$ such Prym differentials that are regular and in fact our
definition depends on an arbitrary point $z_0$, where~\eq{pe-b-cl} has a
simple pole.  However it is not difficult to check that, thanks to the delta
function~\eq{df-cl}, the exponent in~\eq{zm-cl} can be rewritten as
\beqa\label{expO}
\Big\{\ldots\Big\} & =&  
\frac{1}{2}\sum_{\hat\mu,\hat\nu=1}^{g-1} \a_0^{\hat\mu} G {\cal
  S}_{\hat\mu}^{-1} 
\left[ \int\limits_w^{S^{\rm cl}_{\hat\mu}(w)} \!\!\bm{\Omega}_{\hat\nu}(z) -
\frac{1-{\cal S}_{\hat\mu}}{1-{\cal S}_g}
\int\limits_w^{S^{\rm cl}_g(w)} \!\!\bm{\Omega}_{\hat\nu}(z) \right] 
\a_0^{\hat\nu}~,
\\ \label{Omega} &{\rm where} &
\bm{\Omega}_{\hat\nu}(z) = \bm{\zeta}_{\hat\nu}(z)
-\frac {1-{\cal S}_{\hat\nu}}{1-{\cal S}_g}
\bm{\zeta}_g(z)~,~~{\rm with}~~
\hat\nu=1,\ldots,g-1~,
\eeqa
where we have supposed ${\cal S}_g\neq 1$; we need not to worry about the
ordering of the matrices ${\cal S}_\mu$ since they all commute.  In
ref.~\cite{Russo:2003tt} it is shown that in~\eq{expO} the dependence on $z_0$
disappears\footnote {This is also evident from \eq{zeta}.}. Therefore the
$\bm{\Omega}_{\hat\nu}(z)$'s are a set of $g-1$ regular Prym differential and
provide a complete basis for this space. This is the basis that is naturally
selected by the sewing procedure. It is also possible to check that~\eq{expO}
does not depend on the the curve representing the $b_{\hat{\mu}}^{\rm
  cl}$-cycles, {\em i.e.}  connecting $w$ and $S^{\rm cl}_{\hat{\mu}}(w)$.
First, $\bm{\Omega}_{\hat\nu}(z) dz$ is a closed and regular differential so
the exact form of the path is immaterial. Then we can check the independence
of the point $w$, by taking the derivative of the exponent with respect to $w$
and showing that it is zero.  This is more easily done by using the
form~\eq{zm-cl}. Thanks to the periodicities~\eq{perio}. we get
\beq\label{checkw}
\frac{d}{dw} \Big\{\ldots\Big\} =\left\{ \frac{1}{2}
\sum_{\mu,\nu=1}^g \a_0^\mu   
G (1-{\cal S}_\mu^{-1}) 
\bm{\zeta}_\nu(z) \a_0^\nu\right\}~, 
\eeq
where the sums over $\mu$ and $\nu$ are decoupled. Then by using~\eq{orth-pro}
and the presence of the constraint~\eq{df-cl} in the integrand one can see
that~\eq{checkw} vanishes implying that the exponential contribution due to
the zero-modes is independent of~$w$. The independence of~\eq{expO} of $w$
means that the integrals span a closed contour on the {\em branched} Riemann
surface. In terms of the basic cycles of the closed string parameterization
the integral in~\eq{expO} forms the closed contour $b_{\hat{\mu}}^{\rm cl}\,
b_{g}^{\rm cl}\, (b_{\hat{\mu}}^{{\rm cl}})^{-1} (b_{g}^{{\rm
    cl}})^{-1}$. We will depict this path in the open string
parameterization, see Fig.~\ref{dinfty}. 

Thus we can put together the oscillators and the zero-mode contribution and
write the full bosonic partition function of Fig.~\ref{clos-par} for
magnetized bosonic D-branes on a $T^{2d}$
\beqa\nonumber
Z_g^{\rm cl} (F)& =& \left( \prod_{i = 0}^g \sqrt{{\rm
    Det}(1-G^{-1}{\cal F}_i)} 
\right) \int \left[ d Z \right]_g^{\rm c} \, 
\prod_{a=1}^d [{\cal R}^a \left(\vec{\e}^a\right)]^{\rm cl}_g
\sum_m \delta[(1-{\cal S}_\mu){\a}_0^\mu] \\ \label{clchT} &\times &
\exp\left\{\pi\ii \sum_{\hat\mu,\hat\nu=1}^{g-1} \a_0^{\hat\mu} G {\cal
    S}_{\hat\mu}^{-1/2} \bm{D}_{\hat{\mu}\hat{\nu}}{\cal S}_{\hat{\nu}}^{1/2}
  ~\a_0^{\hat\nu}\right\}~,
\eeqa
where the $\a_0$'s are functions of the windings $m$ as dictated by
Eqs.~\eq{nar-lat} and~\eq{Rcl3} and
\beq\label{DA}
\bm{D}_{\hat{\mu}\hat{\nu}} =\frac{1}{2\pi\ii}
{\cal S}^{-1/2}_{\hat{\mu}}
\left[\int\limits_w^{S^{\rm cl}_{\hat\mu}(w)}  -~~\frac{1-{\cal
        S}_{\hat\mu}}{1-{\cal S}_g} \int\limits_w^{S^{\rm cl}_g(w)}
    ~\right]\bm{\Omega}_{\hat\nu}(z) dz~
 {\cal S}_{\hat{\nu}}^{-1/2}~,~~~~~ \hat{\mu},\hat{\nu}=1,2,..,g-1 ~.
\eeq
In the special case of a two-dimensional torus, $T^2$, our~\eq{expO}
is directly connected to the results obtained
in~\cite{Antoniadis:2005sd}. Let us see give some more details to see
how the connection between the two formulations works.

The 1-forms $\bm{\zeta}_\nu(z) dz$ and $\bm{\Omega}_{\hat\nu}(z) dz$ display a
space-time dependence that can be more easily expressed in the basis where all
the monodromy matrices $R_\mu$ are diagonal.  In this eigenvector basis also
$\bm{\zeta}_\nu$ and $\bm{\Omega}_{\hat\nu}$ are diagonal and each entry $
\zeta_\nu^{\pm \vec{\e}^a}$ or $\Omega_{\hat\nu}^{\pm \vec{\e}^a}$ depends on
the eigenvalues $\ex{\pm 2\pi\ii\vec{\e}^a}$ of the corresponding eigenvector.
The $\Omega_{\hat\nu}^{\pm \vec{\e}^a} dz$ are closely related to the 1-forms
$\omega$'s of~\cite{Antoniadis:2005sd}. In fact one can see that their
$\omega_{\hat{\mu}}$ corresponds to our combination
$\ex{-\pi\ii\e_{\hat{\mu}}} \Omega^{\vec{\e}^a}_{\hat\mu}/(2\pi\ii)$.  In
particular one can recover the approximated expression~(A.35)
of~\cite{Antoniadis:2005sd} by rewriting the Prym differential to be inserted
in ~\eq{Omega} in the following form~\cite{Russo:2003tt}\footnote{In the first
  line, ${\sum_\a}^{(\nu)}$ means that in $ T^{\rm cl}_\a$ the last factor on
  the right cannot be a positive or negative power of $ S^{\rm cl}_\nu$; in
  the second line, $a_\nu^\a = \chi_\nu$ if $T^{\rm cl}_\a$ is of the form
  $T^{\rm cl}_\a= T^{\rm cl}_\beta~ (S^{\rm cl}_\nu)^l$ with $l\geq 1$, while
  $a_\nu^\a = \bar{\chi}_\nu$ otherwise. }
\beqa
\zeta_\nu^{\vec{\e}^a} (z) & = & {\sum_\a}^{(\nu)} 
\ex{ 2 \pi \ii (\vec{\e}^a \cdot \vec{N}_\a
+ \e_\nu^a)} \left[\frac{1}{z - T^{\rm cl}_\a (\chi_\nu)} -
\frac{1}{z - T^{\rm cl}_\a(\bar{\chi}_\nu)} \right]
\nonumber \\ & + &
(1 - \ex{ 2 \pi \ii \e_\nu^a}) \sum_\a \ex{ 2 \pi \ii \vec{\e}^a 
\cdot \vec{N}_\a} \left[\frac{1}{z - T^{\rm cl}_\a(z_0)} - \frac{1}{z - 
T^{\rm cl}_\a(a_\nu^\a)} \right]~,
\label{zeta}
\eeqa 
and by taking in the sums only the contribution of $T_\a^{\rm cl}=1$.

In~\cite{Antoniadis:2005sd}, the zero-mode contribution to the twisted
partition function is computed in a IIA setup where D-branes intersect on a
$T^2$. By performing a T-duality we can rewrite the exponential term
in~\eq{clchT} in this language. On a $T^2$ we can write the metric and the $B$
field in the integral basis as follows
\begin{equation}
\label{GB2}
G = \frac{T_2}{U_2}\,
\begin{pmatrix}1 & U_1 \\ U_1 & |U|^2\end{pmatrix}~~~~{\rm and}~~~~
B = 
\begin{pmatrix}\,
0 & - T_1 \\ T_1 & 0
\end{pmatrix}~,
\end{equation}
where $T$ and $U$ are the K\"ahler and complex structure of the
torus. This parameterization is very convenient in the discussion of
T-duality transformations. Indeed, a T-duality along the $x=x^1$ axis
amounts just to the exchange $T \leftrightarrow U$, while a T-duality
along $y=x^2$ corresponds to $T \leftrightarrow -1/U$. For our
computation, we will need two other matrices:
\begin{equation}
\label{F2d}
2\pi\alpha' F_i = 
\begin{pmatrix}
0 & f_i \\ - f_i & 0
\end{pmatrix}~,~~~{\rm and}~~~
{\cal E} = \sqrt{\frac{T_2}{2 U_2}}
\begin{pmatrix}
1 & U \\ 1 & \overline U
\end{pmatrix}~,
\end{equation}
where the magnetic fields $f_i$'s are integers as consequence of~\eq{qc}
and~\eq{commute}, and ${\cal E}$ is the Vielbein matrix transforming the real
integral coordinates into the complex ones that diagonalize the $R_i$'s. In
the complex basis each element of $\bm{D}$ (defined in~\eq{DA}) is a diagonal
space-time matrix. By comparing our conventions and those
of~\cite{Antoniadis:2005sd} we see that its first component
$D_{\hat{\mu}\hat{\nu}}(\e)$ agrees with the definition given in the second
line of~(A.16) of~\cite{Antoniadis:2005sd}. The second component has, of
course, opposite twists and will be indicated by
$D_{\hat{\mu}\hat{\nu}}(-\e)$. If we use~\eq{nar-lat} and~\eq{Rcl3}, we can
re-express the left momentum $\a_0^i$ in terms of the integer windings
\beq\label{v}
{\cal E} \a_0^{i} = \frac {1}{\sqrt{2}}
{\cal E}G^{-1} (G-B-2\pi\a' F_{i}) {m}_{i} = 
\frac {|v_i|}{\sqrt{2}}
R_{0}^{'\,-1/2} {\cal S}_{i}^{'\,-1/2} \frac{|U|}{\sqrt{T_2 U_2}} 
{\cal E} m_i~,~~i=0,1,...g, 
\eeq
where the primed matrices are written in the complex
basis, where they are diagonal, $m_i$ is a column vector with the windings $m^1_i\,,~m^2_i$, and
$|v_i|$ is the T-dual (along $y$) of the same quantity
appearing in~\cite{Antoniadis:2005sd} 
\beq\label{ve}
|v_i|=\sqrt{\frac{U_2}{|U|^2}\frac{|T|^2}{T_2}}\;\frac{|\ii (f_i - T_1) +
  T_2|}{|T|}.  
\eeq
By using~\eq{DA} and~\eq{v} into~\eq{clchT} for the $T^2$ configuration under
study we see that the exponent in~\eq{clchT} can be rewritten as
\beq\label{clchA}
\frac{\pi \ii}{4} (m^1_{\hat{\mu}} + \bar{U} m^2_{\hat{\mu}})
\Big[D_{\hat{\mu}\hat{\nu}}(\e) + D_{\hat{\nu}\hat{\mu}}(-\e) \Big]
(m^1_{\hat{\nu}} + U m^2_{\hat{\nu}}) \frac{|U|^2}{U_2^2}
|v_{\hat{\mu}}||v_{\hat{\nu}}|~.
\eeq
By using the properties of ${D(\pm\e)}$ derived in
Eqs.~(A.14, A.15, A.18) of~\cite{Antoniadis:2005sd},
one can check that $D(\e)+{}^{\rm t}D(-\e)$ coincides with the double
of the matrix $ \tau$ defined in that paper and  that, after a
T-duality $U \leftrightarrow - 1/T$, ~\eq{clchA} agrees with~(5.18)
of~\cite{Antoniadis:2005sd}, apart from an overall factor of $1/2$.
By using the same equations of~\cite{Antoniadis:2005sd}, one can also get the
interesting identity:
\beq\label{C}
D_{\hat{\mu}\hat{\nu}}(\e)-D_{\hat{\nu} \hat{\mu}}(-\e) = 
2\ii \frac{\sin(\pi\e_{\hat{\mu}}) \sin(\pi\e_{\hat{\nu}}-\pi\e_{g})}
{\sin(\pi\e_{g})}~,~~{\rm for}~~\hat{\nu}\geq\hat{\mu}~,
\eeq
that will be checked in Section 4.2 for $g=2$, in the degeneration
limit of the open string channel.

\subsection{Twisted partition function in the open string channel}
\label{op-chan}

\begin{figure}
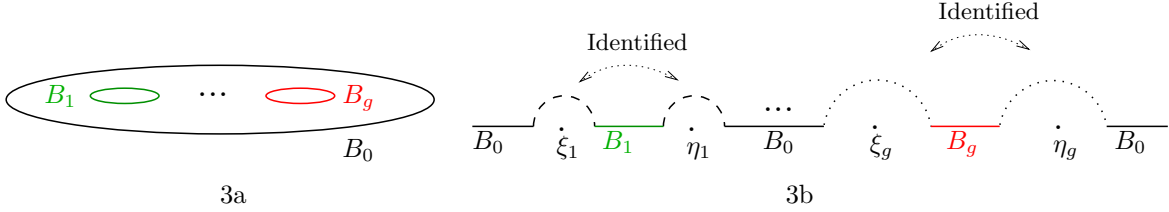

\input nopen-par.pstex_t
\caption{\ref{open-par}a is a space-time representation of the
  same partition function of Fig.~\ref{clos-par} in the open string
  channel; \ref{open-par}b is the corresponding world-sheet surface in
  the Schottky parametrization.\label{open-par}}
\end{figure}

The aim of this section is to perform a modular transformation and to
rewrite the result~\eq{clchT} in the open string channel. Pictorially
this map transforms the world-sheet of Fig.~\ref{clos-par} into the
one of Fig.~\ref{open-par}.
Again the Schottky parameterization of the world-sheet in
Fig.~\ref{open-par}b is completely specified in terms of $g$ $2\times
2$ matrices\footnote{We have endowed the quantities in the closed
string channel with a suffix ``cl''; those without any suffix will
refer to the open string channel, if not otherwise specified.}
\beq\label{Sop}
S_\mu = \left(\begin{array}{cc}
a_\mu &b_\mu \\ c_\mu &d_\mu
\end{array}\right) \equiv \frac{1}{\sqrt{k_\mu}(\eta_\mu-\xi_\mu)}
\left(\begin{array}{cc}
\eta_\mu - k_\mu\xi_\mu& -\eta_\mu \xi_\mu (1-k_\mu)\\
(1-k_\mu)& k_\mu \eta_\mu - \xi_\mu
\end{array}\right)~,
\eeq
where now the fixed points $\eta~,\xi$ (which are the analogue of
$\chi~,\bar\chi$ of~\eq{Scl}) are all real. The open string
world-sheet is the upper half part of the complex plane that is
outside all the circles defined by the $S_\mu$'s as explained
after Eq.~\eq{Scl}. These circles are identified pairwise (see 
Fig.~\ref{open-par}b).

As usual, the modular transformation we are interested in is non-analytic in
the Schottky parameters. This is already manifest in the one loop case where
we have $\ln q=4\pi^2/\ln k$. In order to circumvent this technical problem we
follow the approach used in~\cite{Russo:2003yk,Magnea:2004ai} for the 
twisted partition function in Minkowski space. We first rewrite the products
over the Schottky group in terms of genus~$g$ Theta functions and other
geometrical object such as the Prime Form. This can be done thanks to the
identities derived in~\cite{Russo:2003tt} which are consequence of the
equivalence between fermionic and bosonic theories in two dimensions.  In this
description we can perform explicitly the modular transformation that
exchanges the $a$ and the $b$-cycles\footnote{In the closed string
  parametrization, Fig.~\ref{clos-par}b, we define the $b^{\rm
    cl}_\mu$-cycles to be the segments between $w$ and $S^{\rm cl}_\mu(w)$,
  while the $a^{\rm cl}_\mu$-cycles are the contours around the repulsive
  fixed points $\bar\chi_\mu$ clockwise oriented. In a similar way, in
  the open string parametrization, the $b$-cycles are the segments
  between $w$ and $S_\mu(w)$, while the $a$-cycles are the contours
  around the $\xi_\mu$'s clockwise oriented.}
as follows: $a_\mu^{\rm cl} =b_\mu$ and $b_\mu^{\rm cl} =a_\mu^{-1}$. After
this map we have a world-sheet surface that looks like Fig.~\ref{open-par}a.
For computational convenience we then use once more the identities
of~\cite{Russo:2003tt} to rewrite the result in terms of the open string
Schottky groups (generated by the $S_\mu$'s in~\eq{Sop}). In the non-compact
case, this computation was performed in~\cite{Russo:2003yk,Magnea:2004ai} and
the result for the partition function in the open string channel is
\beq
Z_g(F)|_{\rm uncomp.} = \left[ \prod_{i = 0}^{g} \sqrt{{\rm
 Det}\left(1-G^{-1}{\cal F}_i\right)}\right]
\int\left[d Z \right]_g \, \prod_{a=1}^d
\left[ \ex{- \ii \pi \vec{\e^a} \cdot \tau \cdot
\vec{\e}^a} \; \frac{\det \left(\tau \right)}{\det
\left(T_{\vec{\e}^a} \right)} \; {\cal R}_g \left(
  \vec{\e}^a \cdot \tau \right) \right]~.
\label{effstr}
\eeq
Here $\tau_{\mu\nu}$ is the usual period matrix (written in the open string
channel) and $det$ is the determinant over the ``loop'' indices $\mu\,,~\nu =
1,2,...,g$; $T_{\vec{\e}^a}$ is a twisted generalization of the period matrix
\beqa
& (T_{\vec{\e}^a})_{\hat\nu \hat\mu} = \frac{1}{2 \pi \ii}
\int\limits_{w}^{S_{\hat\nu} (w)} 
\widehat{{\Omega}}^{\vec{\e}^a}_{\hat\mu} (z)
d z ~~,~~{\rm where}~~ 
\widehat{{\Omega}}^{\vec{\e}^a}_{\hat\mu} (z) = 
\left[{\Omega}^{\vec{\e}^a\cdot\tau}_{\hat\mu} (z)
  \ex{\frac{2 \pi \ii}{g - 1} \vec{\e}^a \cdot \vec{\Delta}_z }
\right]
\label{taue} & \\ \nonumber &
  (T_{\vec{\e}^a})_{\hat\nu g}=(T_{\vec{\e}^a})_{g \hat\mu}=0
  ~~,~~{\rm and}~~ 
  (T_{\vec{\e}^a})_{g g}=\frac{ \ex{2\pi\ii(\vec{\e}^a\cdot\tau)_g} - 1}
  {\ex{2\pi\ii \vec{e}^a_g} -1}~~.
\eeqa
We will also use $\widetilde{T}_{\vec{\e}^a}$ to indicate the 
$(g-1)\times (g-1)$ block of ${T}_{\vec{\e}^a}$ which has indices
$\hat{\mu},\hat{\nu}=1,\ldots,g-1$. The Prym differentials
${\Omega}_\mu^{\vec{e}^a\cdot\tau}$ in this equation have the same functional
form of those in~\eq{Omega}, but with the following changes:
\beq\label{cltoop}
\vec{\e}^a ~\to~\vec{\e}^a \cdot \tau~~,~~~
(\bar\chi\,,~\chi)~\to~(\xi,~\eta)  ~~,~~~
S_\mu^{\rm cl} ~\to~ S_\mu  ~.
\eeq
$\vec{\Delta}_z$ is the vector of Riemann constants (or Riemann class)
defined with respect to the base point $z$. When we change the base
point, $\vec{\Delta}$ changes in a simple way determined by the usual
Abelian differentials $\omega$
\beq\label{Dch}
\vec{\Delta}_{(z)} = \vec{\Delta}_{(z_0)} - \frac{g-1}{2\pi\ii}
\int_{z_0}^z \vec{\omega}~.
\eeq
Then it is easy to see that the differentials
$\widehat{{\Omega}}^{\vec{\e}^a}$ in~\eq{taue} are twisted {\em only}
along the $a_\mu$-cycles (of the open string channel) with phases $\ex{-2\pi\ii \e_\mu^a}$. The
identities necessary to rewrite~\eq{clch} as~\eq{effstr} are
Eqs.~(13),~(14) and~(16) of~\cite{Russo:2003yk}. They imply also that
${\rm det}(T_{\vec{\e}^a})$ is invariant under $\vec{\e}^a\to-\vec{\e}^a$ and
can be used to derive the modular transformation of each
single~$\bm{\Omega}_{\hat\nu}(z)$. 

In fact to calculate $Z_g(F)$ in
the compact case we need to express the matrix
$\bm{D}_{\hat{\mu}\hat{\nu}}$ defined in ~\eq{DA} in terms of open
string quantities. From ~\cite{Russo:2003yk}  we get a relation between
the determinants in the closed and open string parameterization and
from there the modular transformation for each
$\bm{\Omega}_{\hat\nu}$:
\beq\label{mod-Omega0}
\overline{\rm det}\Big[\bm{\Omega}_{\hat\nu}(z_{\hat\mu})\Big] = 
\frac{1-{\cal S}_{g}^{\e\cdot\tau}}{1-{\cal S}_g^{\e}}\frac{
\overline{\rm det}\Big[\widehat{\bm{\Omega}}_{\hat\nu}(z_{\hat\mu})\Big]}
{{\det}~ \bm{T}}=
\frac{
\overline{\rm det}\Big[\widehat{\bm{\Omega}}_{\hat\nu}(z_{\hat\mu})\Big]}
{\overline{\det}~ \bm{\widetilde T}}~,
\eeq
where $\widehat{\bm{\Omega}}_{\hat\nu}$,  $~\bm{T}_{\nu \mu}$    and 
$\bm{\widetilde{T}}_{\hat\nu\hat\mu}$ are $2d\times 2d$ 
matrices whose elements in the complex basis (where they are diagonal) 
are the $\widehat{\Omega}_{\hat\nu}^{\pm\vec{\e}^a}$, 
 $~(T_{\pm \vec{\e}^a})_{\nu \mu}$   and
$(\widetilde{T}_{\pm \vec{\e}^a})_{\hat\nu\hat\mu}$ defined in
Eq.~\eq{taue}. 
 The determinants $\overline{det}$ are taken only over the indices
$\hat\mu,~\hat\nu = 1,\ldots, g-1$ and the superscripts for the matrices
${\cal S}_g$ remind that the eigenvalues are either ${\e_g}$ or
${(\e\cdot\tau)_g}$. We can single out each differential
$\bm{\Omega}_{\hat\nu}$ simply by integrating the l.h.s. of~\eq{mod-Omega0} on
the variables $z_{\hat{\mu}}$ with $\hat{\mu}\not= \hat{\nu}$ along the
$a_{\hat{\mu}}^{\rm cl}$-cycles and using the normalization property of the
$\bm{\Omega}_{\hat\nu}$'s: $\oint_{a^{\rm cl}_{\hat\mu}}
\bm{\Omega}_{\hat\nu}= 2\pi\ii \delta_{\hat\mu \hat\nu}$. The r.h.s.
of~\eq{mod-Omega0} is thus integrated along the corresponding
$b_{\hat{\mu}}$-cycles and we get
\beq\label{mod-Omega}
\bm{\Omega}_{\hat\nu}(z) =
\sum_{\hat{\rho}=1}^{g-1} \widehat{\bm{\Omega}}_{\hat\rho}(z)
\bm{\widetilde  T}
^{-1}_{\hat{\rho}\hat{\nu}}
~,~~~~\hat\nu=1,2,\ldots,g-1~.  
\eeq
where the inverse of $\bm{\widetilde T}$ in the last step is over both
the loop and the space-time indices. Notice that this is the natural
generalization of the usual modular property of the Abelian
differentials to the twisted case of Prym differentials, where the
matrix $\bm{\widetilde T}$ plays the role of the period matrix.

Now we can write the generalization of~\eq{effstr} to the case of
toroidal compactification
\beqa \label{opchT}
Z_g (F)& =& \left[ \prod_{i = 0}^{g} \sqrt{{\rm
 Det}\left(1-G^{-1} {\cal F}_i\right)}\right]
\int\left[d Z \right]_g 
\sum_m \delta[(1-{\cal S}_\mu){\a}_0^\mu]
\\ \nonumber &\times&
\exp\!\left\{\!\pi\ii \sum_{\hat\mu,\hat\nu=1}^{g-1} \a_0^{\hat\mu} G {\cal
    S}_{\hat\mu}^{-1/2} \bm{D}_{\hat{\mu}\hat{\nu}}{\cal S}_{\hat{\nu}}^{1/2}
  ~\a_0^{\hat\nu}\right\} \, \prod_{a=1}^d
\left[ \ex{- \ii \pi \vec{\e^a} \cdot \tau \cdot
\vec{\e}^a} \; \frac{\det \left(\tau \right)}{\det
\left(T_{\vec{\e}^a} \right)} \; {\cal R}_g \left(
  \vec{\e}^a \cdot \tau \right) \right]\,,
\eeqa
where  the
matrix $\bm{D}$ of~\eq{DA} is written in terms of the open string variables:
\beq \label{DAo}
 \bm{D}_{\hat{\mu}\hat{\nu}} = 
\frac{1}{2\pi\ii}
{\cal S}^{-1/2}_{\hat{\mu}}
\left[\int_{a_{\hat\mu}^{-1}} \! -~\frac{1-{\cal S}_{\hat\mu}}{1-{\cal S}_g}
\int_{a_g^{-1}} \;\right]
\bm{\Omega}_{\hat\nu}(z) dz~
 {\cal S}_{\hat{\nu}}^{-1/2},~
\eeq
and $\bm{\Omega}_{\hat\nu}$ is given in terms of the open string
channel Prym differentials $ \widehat{\bm{\Omega}}_{\hat\rho}$ by
Eq.~\eq{mod-Omega}.

The complication of this formula lies in the fact that it is not
algebraic since, contrary to what happens in the untwisted case, there
are still various integrals that we are not able to perform in
general. On the other hand it is written in a form that is particular
useful in the study of the degeneration limits as we will now see.

\sect{Yukawa couplings from the partition function}
\label{yuk}

\subsection{The uncompact case}
\label{flat}

\begin{figure}
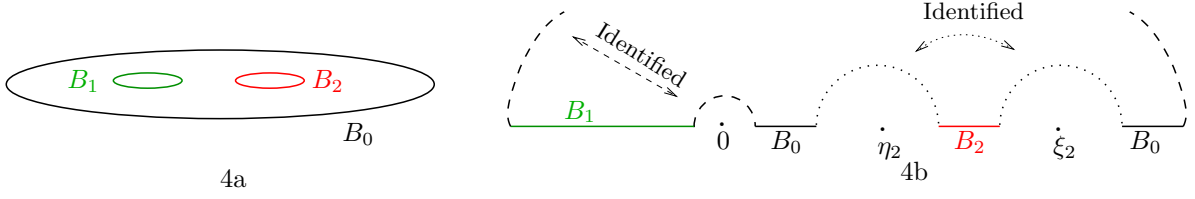

\input open-par.pstex_t
\caption{This figure represents the special case $g=2$ of the partition
  function depicted in Fig.~\ref{open-par}. Notice that for this case we
  choose a slightly different the world-sheet parametrization
  (\ref{open-par2}b) with the respect to the general case: we choose to have
  $\xi_1=\infty$ and $\eta_1=0$ and to swap the order of $\xi_2$ and
  $\eta_2$.\label{open-par2}}
\end{figure}

For $g=2$, the Eq.~\eq{opchT} describes the twisted bosonic partition
function depicted in Fig.~\ref{open-par2}. In the world-sheet
parametrization, we choose to fix $\xi_1=\infty$ and $\eta_1=0$ for
the first pair of disks and $\xi_2=1$ and $\eta_2=\eta$ for the second
pair. Moreover Fig.~\ref{open-par2}b differs from Fig.~\ref{open-par}b
with $g=2$ because we have changed the role of the repulsive and
attractive points of the Schottky generator $S_2$, amounting to send
$S_2$ into $S_2^{-1}$; this has been done to follow usual conventions
for the open string, but has the price that from now on $\e_2$ will be
the opposite of the one used in the previous Sections.

We now will focus on the degeneration limit depicted in
Fig.~\ref{fact}b that is captured by the following region of the
world-sheet moduli space
\beq\label{ftl}
0<k_1 < k_2\ll\eta\ll 1~.
\eeq
The formalism derived in the previous section is particularly suitable
for studying quantitatively the limit~\eq{ftl}, since in this
approximation we need to keep only the first terms in all the Schottky
series or products.  The aim of this section is to focus on the
uncompact case and compute explicitly the leading order contribution
to~\eq{effstr} as function of the world-sheet moduli and the
background fields. Then, as anticipated in the Introduction, we will
obtain by factorization the coupling among three twists fields.
Notice that the limit~\eq{ftl} is similar to that studied
in~\cite{Magnea:2004ai}, where the twisted string partition function
was connected to the field theory Euler-Heisenberg effective action
induced by a charged scalar. The main difference is that now we do not
take any scaling for the $\e_\mu$'s.

As a first step, let us write all the objects entering in~\eq{effstr}
at the first orders in the limit~\eq{ftl}. The Schottky product ${\cal
  R}_2(\e\cdot\tau)$ is just one at the leading order, so we need to
focus only on $T_{\vec{e}^a}$ and $\tau$
\beqa
\label{dete1}
\zeta_1^{\vec{\e} \cdot \tau} (z) & \simeq &
\frac{E_1}{z} + \left[ (1-E_1) \left(\frac{1}{z-z_0}\right) +
(1-E_1) \left(\frac{E_1}{z-k_1 z_0} - \frac{E_1}{z} \right)\right]
~,~~
\\ \nonumber
\zeta_2^{\vec{\e} \cdot \tau} (z) & \simeq & \frac{E_2}{z - \eta} -
\frac{E_2}{z - 1} + (1-E_2) \left(\frac{1}{z-z_0} -
  \frac{1}{z-1}\right)  
\\ \nonumber  & &
+ ~\left[\frac{E_1 E_2}{z - k_1 \eta} - \frac{E_1 E_2}{z - k_1} 
+ (1-E_2) \left(\frac{E_1}{z-k_1 z_0} - \frac{E_1}{z-k_1}\right)\right]\,, 
\\ \nonumber
E_i & = & {\rm e}^{2 \pi {\rm i} \left({\e}_1\tau_{1i}
  +{\e}_2\tau_{2i} \right)}~,~~~ 
~~
\tau \simeq \frac{1}{2\pi\ii}\left(\begin{array}{cc}
\ln k_1 & \ln\eta \\ \ln\eta &\ln k_2
\end{array}\right)~,
\\ \nonumber 
2\pi\ii \vec{\Delta}_z & \simeq  & \left\{-\frac{\ln k_1}{2}
  +\ii\pi - \ln z \;,\;  -\frac{\ln k_2}{2} +\ii\pi
  + \ln\left(\frac{z-1}{z-\eta}
    \frac{z-k_1}{z-k_1\eta}\right)\right\}~.
\eeqa
The first line is obtained from~\eq{zeta} with the
substitution~\eq{cltoop}, where we kept only the identity and $S_1$ in
the Schottky sums since each term in the square parenthesis
in~\eq{zeta} is of order $\prod_\mu k_\mu^{N_\mu}$. Actually at first
sight the terms related to the element $S_1$, are sub-leading in the
limit~\eq{ftl}, but we need to keep them since they yield a finite
contribution to the integral~\eq{taue}.  After we reduced the
Schottky series to the sum of these few terms, we are free to neglect
also the terms that contain an explicit $z_0$ dependence, since we
known that they do not contribute to $\bm{\Omega}$. Now we need to
evaluate the integral~\eq{taue} in the limit~\eq{ftl}, where the
integrand is just a sum of five terms that have branch-cuts in
$\{0,\eta k_1,k_1, \eta, 1\}$. We take the arbitrary point $w$
in~\eq{taue} to coincide with $\eta$. With this choice all the
integrals take the following form
\beq
I(x) \equiv - \int_{k_1 \eta}^\eta \, \frac{d z}{z - x} \left( 
\frac{z - 1}{z - \eta}  \frac{z - k_1}{z - k_1 \eta} \right)^{\e_2}
 \left( \frac{1}{z}\right)^{\e_1}~,
\label{inta}
\eeq
where $x=\{0,\eta k_1,k_1, \eta, 1\}$. Since $z\ll 1$ we can
approximate $z-1$ just with $-1$ and after a change of variables that
brings the region of integration to the interval $[0,1]$, the
integral~\eq{inta} coincides with the integral representation of the
Appell Hypergeometric function $F_1$
\beq\label{appell-def}
\int\limits_0^1 t^{a-1} (1-t)^{c-a-1} (1-t y_1)^{-b_1} (1-t y_2)^{-b_2} dt =
\frac{\Gamma(a)\Gamma(c-a)}{\Gamma(c)} F_1(a;b_1,b_2;c;y_1,y_2)\,.
\eeq
At this point we have to extract the leading order contribution in the
$F_1$'s. This is a somewhat delicate step because in the limit~\eq{ftl} both
$y_i$ in~\eq{appell-def} tend to one, so the standard series representation of
$F_1$ is not useful. However the two $y_i$'s scale in an ordered fashion. In
fact from~\eq{inta} we see that we can pose $y_1= (1-k_1)/(1-k_1/\eta)$ and
$y_2 = (1-k_1)$ so, in the limit~\eq{ftl}, we have $|1-y_2|
\ll |1-y_1|$. In this
case we can use the following approximation
\beqa
F_1(a;b_1,b_2;c;y_1,y_2) \!\!\!& \sim &\!\!\! \frac{B(c-b_1-b_2-a,a)}
{B(a,c-a)} + (1-y_1)^{-b_1} (1-y_2)^{c-b_2-a}
\frac{\Gamma(c)\Gamma(a+b_2-c)}{\Gamma(a)\Gamma(b_2)} 
\nonumber\\ & + & \label{app-app}
(1-y_1)^{c-b_1-b_2-a}
\frac{\Gamma(c)\Gamma(c-b_2-a)
  \Gamma(b_1+b_2+a-c)}{\Gamma(a)\Gamma(c-a)\Gamma(b_1)}~,
\eeqa
where $B(a,b)$ is the Euler beta function. It is convenient to
introduce the more symmetric variables~\cite{DiVecchia:1996kf}
\beq\label{schw}
q_1 = k_1/\eta=\ex{-t_1/\a'}~~,~~~~q_2 = k_2/\eta=\ex{-t_2/\a'}~~,~~~{\rm
  and}~~~q_3 = \eta=\ex{-t_3/\a'}~,
\eeq
where the $t_i$ are directly related to the Schwinger proper times for
the propagators depicted in Fig.~\ref{fact}c. In this variable, the
degeneration limit becomes $q_1 < q_2< q_3\ll 1$. By using
Eqs.~\eq{app-app} and~\eq{inta} into Eqs.~\eq{dete1} and~\eq{taue} and
also~\eq{ggs}, we get a very symmetric expression
\beqa\label{taue-lo1}
{\det \left(T_{\vec{\e}^a} \right)} & \simeq & -\frac{1}{4\pi^2}
\Bigg\{\Big[\Gamma(-\e_1^a)\Gamma(-\e_2^a)\Gamma(\e_1^a+\e_2^a)
q_1^{\e^a_1/2} q_2^{\e^a_2/2} q_3^{-\e^a_1/2-\e^a_2/2}
\\ \nonumber & \times &
\left(\e^a_1 q_1^{-\e^a_1} + \e^a_2 q_2^{-\e^a_2} - (\e^a_1+\e^a_2)
  q_3^{\e^a_1+\e^a_2} \right) \Big] + 
\Big[ \e^a_i~\leftrightarrow\, -\e^a_i\Big] 
\Bigg\}~,
\eeqa
where the last square parenthesis means that we have to sum a
contribution where $\e_1\,,~\e_2$ have everywhere the opposite sign.
From now on we choose both $\e$'s to be positive, then the dominant
term in the above expression is
\beq\label{taue-lo2}
{\det \left(T_{\vec{\e}^a} \right)} \simeq \frac{1}{4\pi^2}
\Big[\Gamma(\e^a_1)\Gamma(\e^a_2)\Gamma(1-\e^a_1-\e^a_2)
q_1^{-\e^a_1/2} q_2^{-\e^a_2/2} q_3^{-\e^a_1/2-\e^a_2/2}
\Big] + \ldots~,
\eeq
where the terms neglected do not contribute to the correlator among
three twist fields $\sigma_\e$ but are relevant for the couplings
with excited twist fields. Let us know pose $\e_3=1-\e_1-\e_2$ so that
the three parameters $\e_i$ satisfy the relation that if often
considered in the twist field couplings $\sum_{i=1}^3 \e_i = 1$. By
using~\eq{taue-lo2} and~\eq{dete1} into the integrand of
Eq.~\eq{effstr} we have
\beqa
Z_2 (F)|_{\rm uncomp.} &\simeq & \prod_{i = 0}^{2} \left[{\rm
 Det}\left(1-G^{-1} {\cal F}_i\right) \left(\prod_{a=1}^d
\sin\pi\e_{i+1}^a\right)\right]^{1/2}  
\int \prod_{i=1}^3 
\left[ q_i^{\e_i-\e_i^2} 
\frac{\Gamma(1-\e_i)}{\Gamma(\e_i)} \right]^{1/2}
\nonumber \\ \label{effstr-l1} & \times &
\left(\frac{1}{\ln q_1 \ln q_2 + \ln q_1 \ln q_3 + \ln q_2 \ln q_3}
\right)^{13-d}
\prod_{i=1}^3\frac{d q_i}{q_i^2}~,
\eeqa
where we used the approximate expression for $[d
Z]_{g=2}$~\cite{DiVecchia:1996kf} and the identity
\beq\label{ggs}
\Gamma(z) \Gamma(1-z) = \frac{\pi}{\sin\pi z}~.
\eeq
Now we are in the position of performing the last step and factorize
the degenerate surface in Fig.~\ref{fact}b into its constituents, as
depicted in Fig.~\ref{fact}c. In this computation we will ignore all 
the terms that do not depend on the world-sheet
moduli. In the next section we will see that these normalizations
combine nicely with others terms originating from the zero-mode
contributions. The factors of $q_i^{\e_i/2-\e_i^2/2}$
in~\eq{effstr-l1} and the measure $dq_i/q_i^2$ present in the
untwisted $[d Z]_{g=2}$ take a simple form in the variables~\eq{schw}
and yield the measure $dt_i \exp{\left[-M_{\e_i }^2 t_i\right]}$, where 
$M_{\e_i}$ is exactly the mass of the twisted tachyon~\eq{twi-tac} for
each $\e_i$. This term together with the leading order of the period
matrix determinant have a very simple space-time interpretation: they
are just a representation of the three massive propagators, with the
momentum conservation present in the vertices enforced
\beq\label{fact-prop}
\int \left(\prod_{i=1}^3 dt_i\right) 
  \frac{\sum_i \exp{\left[-M_{\e_i }^2 t_i\right]}}{(t_1 t_2 + t_1
    t_3+ t_2 t_3)^{13-d}} = 
\int  \left(\prod_{i=1}^3  d^{26-2d} p_i \frac{1}{p_i^2 + M_{\e_i }^2}
\right)  \delta(\sum_i p_i) ~.
\eeq
Thus if we compare Eq.~\eq{effstr-l1} and Fig.~\ref{fact}c we can
obtain an explicit expression for the square of the string vertex
containing three twists fields. In formulae (and, for the moment,
neglecting the normalization) we get
\beq\label{u-twist}
\langle \sigma_{\e_1}\sigma_{\e_2}\sigma_{\e_3} \rangle
= \prod_{i=1}^3 \left[\frac{\Gamma(1-\e_i)}{\Gamma(\e_i)}
  \right]^{1/4}~.
\eeq
This result agrees with previous
results~\cite{Burwick:1990tu,Cvetic:2003ch}. 

\subsection{Twist field couplings on a $T^2$}
\label{compact}

The new ingredient in the toroidal partition function~\eq{opchT} is the
exponential due to the zero-mode contribution. This exponential depends both
on the twisted period matrix $T_{\vec{\e}^a}$ we have already encountered in
the previous section and on new integrals of the Prym differentials. In the
two loop case~\eq{DAo} simplifies since in this case there is only one regular
Prym differential and one element in the matrix $\bm{D}_{11} \equiv \bm{D}$
\beq\label{2lexp}
\bm{D} = \frac{1}{2\pi\ii}
\frac{{\cal S}_1^{-1}}{(1-{\cal S}_2)^2}
\left[(1-{\cal S}_2)\int_{a_1^{-1}} \! -\; (1-{\cal S}_1)
\int_{a_2^{-1}} \;\right]
\frac{(1-{\cal S}_{2}^{\e\cdot\tau})\widehat{\bm{\Omega}}(z)}
{{\det}~ \bm{T}}
 dz ~; 
\eeq
moreover in~\eq{opchT}~ $\a_0\equiv \a_0^{\mu=1}$ is the left moving
momentum~\eq{nar-lat} running in the first loop. Each integral over
the $a_1$ and the $a_2$ cycle does {\em not} follow a close path
because of the monodromies of $\widehat{\bm{\Omega}}$. However, from
our analysis in the closed string channel~\eq{checkw}, we know that the
combination of integrals in the square parenthesis is independent of
the starting point. Thus this combination should represent a closed
path on the branched Riemann surface.
\begin{figure}
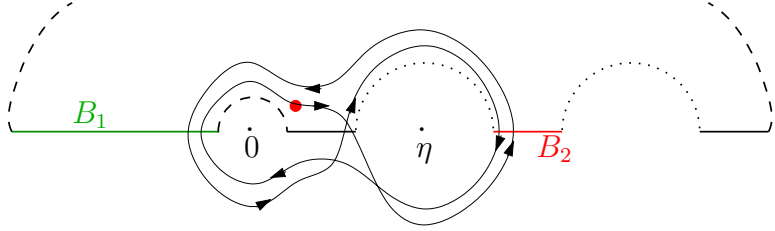

\begin{center}
\input dinfty.pstex_t
\end{center}
\caption{This represents the path of integration ${\cal C}$ resulting
  from the combination of the integrals in the square parenthesis
  of~\eq{2lexp}. Starting from the dot and following the arrows one
  can check that this is a closed circuit on the branched Riemann
  surface. The $\e_1$-cut ($\e_2$-cut) are along the border $B_1$
  ($B_2$). \label{dinfty}}
\end{figure}
In fact, if we follow the cycle $a_2^{-1}$, we have to through the cut $B_2$,
and the integrand is multiplied by ${\cal S}_2$. So it is smoothly connected
to the last integral ${\cal S}_2 \int_{a_1^{-1}}$ and picks a phase ${\cal
  S}_1$ when passes through the cut $B_1$ following $a_1^{-1}$. Then the
result is smoothly connected to the second integral after we transformed the
anti-clockwise orientation of ${a}_2^{-1}$ into the clockwise one
(${a}_2$). In presence of the cut along $B_2$ we have ${\cal S}_1
\int_{{a}_2} = -{\cal S}_1 {\cal S}_2 \int_{{a}_2^{-1}}$. Repeating the same
steps we see that the third and last integral brings back the integrand on the
sheet where we started from, thus showing that the path of Fig.~\ref{dinfty}
is closed. At leading order in the limit~\eq{ftl}, the integrand
$\widehat{\bm{\Omega}}$ is defined through the expansions in~\eq{dete1}, where
this time we can neglect all the terms in the square parenthesis. In fact,
contrary to what happened in~\eq{inta}, the path of integration depicted in
Fig.~\ref{dinfty} never comes close to the points $0$ and $\eta=q_3$. After we
truncated the integrand in~\eq{2lexp} at the first order we are free to deform
the contour to the real axis obtaining 
\beq\label{414}
\left[(1-{\cal S}_1) \int_{a_2^{-1}} \! -\; (1-{\cal S}_2)\int_{a_1^{-1}}
  \;\right] = 
(1-{\cal S}_1) (1-{\cal S}_2) \int_0^{q_3}~.
\eeq
By using this relation in~\eq{2lexp} we have
\beq\label{Do}
\bm{D} = -\frac{1}{2\pi\ii} \frac{1-{\cal S}_1^{-1}}{1-{\cal S}_2}
\frac{\bm{f}}{{\det}~ \bm{T}}~,
\eeq
where the function $\bm{f}$ is the integral between $0$ and $q_3$ of
the Prym differential $(1-{\cal S}_{2}^{\e\cdot\tau})
\widehat{\bm{\Omega}}$ approximated to the first
order. From~\eq{dete1} and using that $z\ll 1$, we have
\beq
f(\e) \sim q_1^{-\frac{\e_1}{2}} q_2^{-\frac{\e_2}{2}}
q_3^{\frac{\e_1+\e_2}{2}} \int_0^{q_3} \left[ 
\frac{q_1^{\e_1} q_3^{\e_1+\e_2} -1}{z-{q_3}}~ q_2^{\e_2}
- \frac{q_2^{\e_2} q_3^{\e_1+\e_2} -1}{z}~q_1^{\e_1} \right] 
\frac{\ex{\pi\ii(\e_1+\e_2)} dz}{({q_3}-z)^{\e_2} z^{\e_1}}\,,
\eeq
where $f(\e)$ indicates the first component of the space-time matrix $\bm{f}$
that is diagonal in the complex basis. The other component is, of course,
$f({-\e})$ where the signs of all twists are reversed. After the change of
variable $y=z/{q_3}$, this integral reduces to the integral definition of the
Euler Beta. Then, by using~\eq{ggs}, we obtain a simple expression for
$\bm{D}$, valid in the limit~\eq{ftl}
\beqa\nonumber
\bm{D} 
& \sim & \Bigg\{
{\frac{\sin(\pi\e_1+\pi\e_2)\sin{(\pi \e_1)}}{\pi\sin{(\pi\e_2)}}} 
q_1^{\e_1/2} q_2^{\e_2/2} q_3^{-\e_1/2-\e_2/2} \Gamma(-\e_1) \Gamma(-\e_2)
\Gamma(\e_1+\e_2) \\ \label{bmf}&\times & \frac{\left(\e_1 q_1^{-\e_1}+ \e_2
  q_2^{-\e_2} -(\e_1+\e_2) q_3^{\e_1+\e_2}\right)}{-2\pi\ii \det T_{\vec{\e}}}
\;,\; \e_i \leftrightarrow
-\e_i \Bigg\}~.  \eeqa
By using~\eq{taue-lo1}, one can check that 
\beq\label{C'}
D(\e)-D(-\e) = - 
2\ii \frac{\sin(\pi\e_1) \sin(\pi\e_1+\pi\e_2)}
{\sin(\pi\e_{2})}~,
\eeq
which is consistent with the general property~\eq{C}\footnote{Remember that
  here $\e_2$ is the opposite of the one used in the previous Section, as
  discussed at the beginning of Sect.~4.1.}.

The first order of the $q_i\to 0$ limit describes the factorization depicted
in Fig.~\ref{fact}c. We can now use the result we have and check that the
leading term in the exponential~\eq{2lexp} does not depend on the $q_i$'s. By
using also~\eq{v} and~\eq{ve} we obtain
\beq\label{2lexp3} 
{\rm Eq.~\eq{2lexp}} \sim \exp\!\left\{ \!- \frac{\pi}{2}
(m^1 \!\!+ \!\bar{U} m^2)\!\left[
\frac{\sin(\pi\e_{1}) \sin(\pi\e_{1}\!+\!\pi\e_{2})}{\sin(\pi\e_{2})}
\;\frac{|\ii (f_1 \!- T_1) + T_2|^2}{T_2 U_2} \right]
\!(m^1\!\! +\! U m^2)\!
\right\}.
\eeq
In the $T^2$ case under analysis, it is easy to relate the eigenvalues of
${\cal S}_\mu$ and the phases $\e_\mu$
\beq\label{calse}
\ex{2\pi\ii\e_2} = \frac{T-f_0}{\bar{T}-f_0}
\frac{\bar{T}-f_2}{{T}-f_2}
~~,~~~
\ex{2\pi\ii\e_1} = \frac{T-f_1}{\bar{T}-f_1}
\frac{\bar{T}-f_0}{{T}-f_0}~.
\eeq
From this one can check that in the square parenthesis in~\eq{2lexp3} all
terms containing the K\"ahler moduli $T_1$ and $T_2$ disappear and one
has
\beq\label{noT}
\left[ \frac{\sin(\pi\e_{1}) \sin(\pi\e_{1}+\pi\e_{2})}{\sin(\pi\e_{2})}
  \;\frac{|\ii (f_1 - T_1) + T_2|^2}{T_2} \right]
= \frac{|f_1-f_0| |f_1-f_2|}{|f_2-f_0|}~.
\eeq
Thus at the leading order, the open string channel partition
function~\eq{opchT} becomes
\beqa\nonumber
Z_2 (F) &\simeq & \sqrt{G} \;\ex{\phi_{10}}
\prod_{i = 0}^{2} \left[{\rm
 Det}\left(1-G^{-1} {\cal F}_i\right) \left(\prod_{a=1}^d
\sin\pi\e_{i+1}^a\right)\right]^{1/2}  
\sum_m \delta[(1-{\cal S}_\mu){\a}_0^\mu] 
\\  \label{effstr-c1}  & \times & 
\exp\left\{ -\frac{\pi}{2}(m^1 \!+ \!\bar{U} m^2) 
\frac{|f_1-f_0| |f_1-f_2|}{|f_2-f_0|} \frac{1}{U_2}
(m^1 \!+\! U m^2)\right\}
\\ \nonumber & \times &
\prod_{i=1}^3 
\left[\frac{\Gamma(1-\e_i)}{\Gamma(\e_i)} \right]^{1/2}
\int \left(\prod_{i=1}^3  d^{26-2d} p_i \frac{1}{p_i^2 + M_{\e_i }^2}
\right)  \delta(\sum_i p_i)~,
\eeqa
where we wrote explicitly also the usual $F_i$-independent normalizations
contained in $[dZ]_2$: the volume ($\sqrt{G}$) and the dilaton factors
($\ex{\phi_{10}}$). It is useful to rewrite the latter in terms of the
effective $d$-dimensional dilaton according to $\ex{2\phi_{10}} =\sqrt{G}
\ex{2\phi_d}$. Then the product over $\sin\pi\e_i$ can be written 
as follows
\beq\label{rwsin}
\prod_{i=1}^3\prod_{a=1}^d \sin^2\pi\e_i^a = \prod_{i=1}^3 {\rm
  Det}\left(\frac{1-{\cal S}_i}{2}\right) = \prod_{i=0}^2
 {\rm Det}\left(\frac{R_{i+1}-R_{i}}{2}\right)~,
\eeq
where in the last step we used the fact the the reflection matrices
$R_i$ are even dimensional and have unit determinant (we also take a
cyclic convention $R_3\equiv R_0$). By using~\eq{commute} and~\eq{rwsin}
into~\eq{effstr-l1}, we see that this determinant combines nicely
with the Born-Infeld normalization yielding
\beq\label{bif}
\prod_{i = 0}^{2} \left[{\rm Det}\left(1-G^{-1} {\cal
        F}_i\right) \prod_{a=1}^d \sin\pi\e_{i+1}^a\right]^{1/2}  
=  \Big( {\rm Det}~ G \Big)^{-3/4} \prod_{i = 0}^{2}\Big[ {\rm
    Det}\left(F_{i+1}-F_i \right)\Big]^{1/4}~.
\eeq
Let us now analyze the second line of~\eq{effstr-c1} and, for sake of
simplicity, suppose that all $(f_i-f_j)$'s which are integers as consequence
of ~\eq{qc}--\eq{commute}, are coprime. Then the Kronecker delta
implies that 
\beq
m_1^M I_{10} + m_2^M I_{20} =0~~\Rightarrow~~~~
m_1^M \equiv m^M = I_{20} h^M~,
\eeq
where $I_{\mu 0}=f_\mu -f_0 = p_\mu-p_0$, $h^M \in Z$ and $M=1,2$. So the
exponential terms form a 2-dimensional Theta function whose modular parameter
contains the complex structure and the product ${\cal I} = |I_{10} I_{02}
I_{21}| $; notice that ${\cal I}= 2 {\cal I'}$ is even because it is the
absolute value of the product of three integers summing to zero. At
this point, we rewrite 
the sum on $h^1$ by performing a Poisson
resummation
\beq\label{poisson}
\sum_{h^1=-\infty}^{\infty} \ex{-\pi A (h^1)^2+2\pi h^1 A s} = 
\frac{1}{\sqrt{A}} \ex{\pi A s^2} \sum_{{h}^1=-\infty}^\infty 
\ex{-\pi ({h}^1)^2/A -2\pi\ii {h}^1 s}~.
\eeq
The second line of~\eq{effstr-c1} becomes
\beq\label{i1}
\sqrt{\frac{U_2}{{\cal I'}}} \sum_{h^1,{h}^2}
\exp\left[-\frac{\pi({h}^1)^2 U_2}{{\cal I'}}+ 2 \pi\ii h^1 {h}^2 U_1 -\pi
  {\cal I'} U_2  (h^2)^2 \right]~. 
\eeq
In this form we can decouple the two sums. However, it is first convenient to
break the sum over ${h}^1$ in ${\cal I'}$ sums over integers of the form
${\cal I'} \tilde{h}^1+ \ell'$, with $\ell'=0,1,..,{\cal I}'-1 $. Then we can
introduce an other set of integers $r,s$: $\tilde{h}^1-{h}^2 = r$ and
$\tilde{h}^1+{h}^2 = s$, which are however constrained to have the same
parity. Therefore the sum over ${r,s}$ breaks into two parts, one over
$r=2k,s=2l$ and the other over $r=2k+1,s=2l+1$. This second sum can be
obtained from the first one simply by replacing $\ell'$ with ${\cal
  I'}+\ell'$. Then we can finally rewrite~\eq{i1} in terms of 1-dimensional
Theta functions\footnote{Notice that now the summation index $\ell$ runs from
  $0$ to ${\cal I}- 1$.}:
\beq\label{i2}
\sqrt{\frac{2 U_2}{{\cal I}}} \sum_{\ell=0}^{{\cal I}-1}\sum_{l,k}
\exp\left[\pi \ii {\cal I} \left(l+\frac{\ell}{\cal I}\right)^2 U
\right]
\exp\left[-\pi \ii{\cal I} \left(k+\frac{\ell}{\cal I}\right)^2
  \bar{U} \right]~.
\eeq
Thus we see that in the limit~\eq{ftl} the full twisted partition factorizes
into a {\em sum} of diagrams like the one depicted in Fig.~\ref{fact}c, and
each diagram is labeled by a different $\ell$. However this sum was expected
and is due to the finite dimensional degeneracy of each twisted string state,
as was quickly mentioned in Section~\ref{conv} in relations to the zero-modes.
From a lower dimensional point of view these states are seen as different
particles of the same family, since they have the same quantum numbers. Since
for each twisted string this degeneracy is exactly $|I_{ij}|$, the number of
possible three point functions is ${\cal I}$ in agreement with our
interpretation of~\eq{i2}. Thus we can focus on each value of $\ell$
separately and derive from~\eq{effstr-c1} the expression for the 3-twist field
correlator on a $T^2$
\beq\label{yu}
\langle \sigma_{\e_1}\sigma_{\e_2}\sigma_{\e_3} \rangle_\ell = 
\left[\prod_{i=1}^3\frac{\Gamma(1-\e_i)}{\Gamma(\e_i)}\right]^{1/4}
\ex{\phi_d/2}  (2 U_2)^{1/4}
\sum_{l=-\infty}^\infty
\exp\left\{\pi\ii{\cal I}\left(l+\frac{\ell}{\cal I}\right)^2 U\right\}~,
\eeq
where the overall normalization assumes a simple form since $F$-dependent part
of~\eq{bif} exactly cancels the factor of $\sqrt{\cal I}$ in~\eq{i2}. For
small $\e$ we reproduce the result of~\cite{Cremades:2004wa}, including the
U-dependent normalization. If we exchange $T$ and $U$ in~\eq{yu}, we pass to
the T-dual setup of D-branes at angle and find agreement with the results
of~\cite{Cremades:2003qj,Cvetic:2003ch}.
In~\cite{Cremades:2003qj}, the amplitudes corresponding to different families
were labeled by three integers $i,j,k$, but it can be checked that their set
of Yukawa couplings corresponds exactly to our~\eq{yu}.

\vspace{1cm}

\noindent {\large {\bf Acknowledgments}}

\vspace{3mm}
\noindent This work is partially supported by the European Community's
Human Potential under contract MRTN-CT-2004-005104 and
MRTN-CT-2004-512194, 
and by the Italian MIUR under contract PRIN 2005023102. We wish to
thank Dario Du\`o and Lorenzo Magnea for collaboration on related
topics and for helpful discussions.

\vspace{1cm}

\appendix

\sect{The twisted sewing on $T^{2d}$}
\label{appa}

In the operator formalism all the basic building blocks like string
vertices and propagators are written in terms of the Schottky group.
In both the closed and the open string description this group is
generated by $2\times 2$ matrices that act on the complex plane as
follows 
\beq\label{g-pro}
\gamma(z) = \left(\begin{array}{cc}
a &b \\ c &d
\end{array}\right) 
\left(\begin{array}{c}
z \\ 1
\end{array}\right)  = 
\frac{a z +b}{ c z +d}~~,~~~{\rm with}~~~
ad-bc=1\,.
\eeq
In the string vertices the bosonic oscillators $a_n$ or $\tilde a_n$
are naturally contracted with the following infinite dimensional
representation of the Schottky group (here $n,m\not=0$)
\beqa\label{D}
D_{nm}(\gamma) =\frac{1}{m!} \sqrt{\frac{m}{n}}
\partial_z^m\gamma^n\Big|_{z=0}
~~&,&~~~~
D_{00} = -\ln d~~,
\\ \nonumber
D_{n0} = \frac{1}{\sqrt{n}} \gamma^n(0) 
~~&,&~~~~
D_{0m} = \frac{\sqrt{m}}{2 m!} \partial^{m}_z \ln\gamma'\Big|_{z=0}
=  \frac{1}{\sqrt{m}} \left(\frac{1}{\gamma^{-1}(\infty)}\right)^m
~.
\eeqa
As already noticed in the main text, this is not a true representation
when the zero modes are included. In fact the product law is 
\beq\label{D-rep}
D_{nm}(\gamma_{1}\gamma_{2})=
\sum_{l=1}^{\infty}D_{nl}(\gamma_{1})D_{lm}(\gamma_{2})+
D_{n0}(\gamma_{1})\delta_{m0}+D_{0m}(\gamma_{2})\delta_{n0}~.
\eeq
This is the source of some complication in the zero-mode computation of the
twisted planar partition function and in this Appendix we want to give some
details on the technical steps leading to~\eq{zm-cl}.

The closed string vertex describing the emission of $g+1$ off-shell
strings can be written in terms of $g+1$ transformations~\eq{g-pro}
$V_i$ 
\beqa \label{close-vert}
{\bf V}_{g+1}&\sim&
\prod_{i=0}^g\left[\sum_{n^i,w^i}\langle n^i,w^i;0| \right]
\exp\!\!{\left[-\!\!\!\sum_{i< j=0}^g\sum_{k,l=0}^\infty \!a_k^{i} 
D_{kl}(\Gamma V^{-1}_i V_j) G a_l^{j} \right]}
\\ &\times& \nonumber
\exp\!\!{ \left[-
\!\!\!\sum_{i< j=0}^g\sum_{k,l=0}^\infty\!{\tilde a}_k^{i}
D_{kl}(\Gamma \bar V^{-1}_i\bar V_j) G {\tilde a}_l^{j} \right]}
\delta\Big({\sum\limits_{i=0}^g a_0^{i}}\Big)\,
\delta\Big({\sum\limits_{j=0}^g \tilde a_0^{j}}\Big)
\;,
\eeqa
where the $\delta$'s should be interpreted as Kronecker delta's and we
neglected the overall normalization. Here the $V_i(z)$'s are local coordinates
around each puncture which is placed at $V_i(0)=z_i$, $\Gamma$ is the
inversion $\Gamma(z) = 1/z$, and since we are working in the parameterization
of Fig~\ref{clos-par} the right-moving coordinates are just the complex
conjugate of the left moving ones. Of course when~\eq{close-vert} is saturated
with $g+1$ on-shell states $V_{g+1}$ in order to compute a tree-level
amplitude the dependence on the local coordinates disappears. However our aim
is to saturate all legs of~\eq{close-vert} with boundary states; in this case
the $V_i$'s play a role, because they yield the generators of the Schottky
group~\eq{Scl} that defines the surface of Fig.~\ref{clos-par}. Also the
boundary states are written in terms of the matrices $D_{nm}$
\beqa \label{general}
&& |B_i\rangle \sim 
 \sqrt{{\rm Det}(1-G^{-1}{\cal F}_i)} \int_0^1 \frac{dx}{x(1-x)} 
\delta^d(\tilde{a}_0+R_i a_0) |B_i(x)\rangle~,
\\ \nonumber &&
|B_i(x)\rangle = \exp{\left[-\sum_{n,m =0}^\infty 
\tilde{a}_n^\dagger
D_{nm}(P(x))\, G\, R_i {a}_m^\dagger\right]}
{\rm e}^{2\pi\ii \sqrt{\a'} (A_i)_M w_i^M}
\left(\sum_{n,w} |n,w;0\rangle\right)~,
\eeqa
where again we neglected all $F_i$ independent normalizations, which can be
found in~\cite{Frau:1997mq} or restored at the end by requiring that the
partition function is proportional to the total volume and the appropriate
power of the string coupling. The phase depending on the Wilson line $A_i$ is
just the rewriting in terms of the winding modes of the usual term $exp\{\ii
A_M\oint dx^M\}$ appearing in the open string action for constant gauge
fields.  With a small abuse of notation we indicate with $|B_i\rangle$ both
the standard boundary state~\eq{Rcl2} whose effect is just to identify left
and right moving modes and the boundary state~\eq{general} that contains also
a string propagator.  This propagator is contained in the transformation
$P(x)$, which is also of the type~\eq{g-pro} and depends on a single real
variable $x$. As we will see the specific form chosen for $P(x)$ and the
$V_i$'s is not relevant for our purposes.

The insertion of the first boundary states $\ket{B_0}$ has the effect of
mixing left and the right moving modes.  This can be easily seen for the
non-zero mode sector, where the boundary state simply transforms the $\tilde
a^0_m$ into creation operators $a_m^{0 \dagger}$. When the scalar product over
the $i=0$ Hilbert space is computed, the exponential factors at the left
containing the destruction modes $a_m^{0}$ are glued together with the
exponentials at the right containing the creation modes $a_m^{0 \dagger}$.
The zero-mode sector has to be treated separately and by using the momentum
and winding conservations present in~\eq{close-vert} we can re-express $a_0^0$
in terms of the $a_0^\mu$, with $\mu=1,\ldots,g$. Then by using the product
law~\eq{D-rep} and the properties summarized in Appendix~B
of~\cite{DiVecchia:1988cy} we get a vertex describing the emission of $g$
closed strings from a disk represented by the boundary state $\ket{B_0}$
\beqa\nonumber
{\bf V}_{g;1}&\sim&
\prod_{i=1}^g\left[\sum_{n^i,w^i}\langle n^i,w^i;0| \right]
\delta\left[\sum_{\mu=1}^g\left(a_0^\mu+R_0^{-1} 
\tilde{a}_0^\mu\right)\right] 
\exp\!\!{\left[-\!\!\!\sum_{\mu,\nu=1}^g 
\!a_k^{\mu} D_{kl}(U_\mu \bar{V}_\nu) G R_0^{-1} \tilde{a}_l^{\nu} \right]}
\\ &\times& \label{vg1}
\!\exp\!\!{\left[-\!\!\!\sum_{\mu< \nu=1}^g 
\!a_k^{\mu} D_{kl}(U_\mu V_\nu) G a_l^{\nu} \right]}
\exp\!\!{ \left[-\!\!\!\sum_{\mu< \nu=1}^g 
\!{\tilde a}_k^{\mu} D_{kl}(\bar U_\mu\bar V_\nu) G {\tilde a}_l^{\nu}
\right]} \,,
\eeqa
where we have understood the sums over the indices $k,l$ from $0$ to $\infty$
and followed the notation of~\cite{DiVecchia:1988cy}, where $U =\Gamma V^{-1}$.
We also used the freedom to redefine the local coordinates $V_\mu$ by a
similarity transformation to eliminate any dependence on $V_0$.

At this point we can insert the remaining $g$ boundary states in~\eq{vg1} and
use again the properties of Canonical Forms~\cite{DiVecchia:1988cy} and
Eq.~\eq{D-rep} to rewrite the exponentials in a normal ordered way
\beqa\nonumber
{\bf V}_{g+1;0}&\sim&
\prod_{i=1}^g\left[\sum_{n^i,w^i}\langle n^i,w^i;0| \right]
\delta\!\!\left[\sum_{\mu=1}^g\left(1-{\cal S}_\mu\right) 
{a}_0^\mu\right] 
\exp\!\!{\left[-\frac 12\!\!\!\sum_{\mu\not= \nu=1}^g 
\!a_k^{\mu} D_{kl}(U_\mu V_\nu) G a_l^{\nu} \right]}
\\ &\times& \label{vg2}
:\exp\!\!{\left[\sum_{\mu,\nu=1}^g 
\!a_k^{\mu} D_{kl}(U_\mu \tilde{\bar{V}}_\nu) G {\cal S}_\nu {a}_l^{\nu
  \dagger} \right]}:
\\ &\times& \nonumber
\exp\!\!{ \left[-\frac 12\!\!\!\sum_{\mu\not= \nu=1}^g 
\!{a}_k^{\mu \dagger} D_{kl}(\tilde{\bar{U}}_\mu\tilde{\bar{V}}_\nu) G 
  {\cal S}_\mu^{-1} {\cal S}_\nu {a}_l^{\nu \dagger} \right]} 
\prod_{i=1}^g\left[\sum_{n^i,w^i}\ket{n^i,w^i;0} \right]
\,,
\eeqa
where the tilded local coordinates are related to the original ones by
$\tilde{V} = V P$. Then we need to compute the scalar products over the
oscillator modes of all Hilbert spaces. This can be done more easily by
inserting the identity operator written in terms of coherent states between
the creation and the destruction modes~\cite{DiVecchia:1988cy}. This has the
effect of transforming the scalar product over the non-zero modes into a
Gaussian integral which yields the following contribution to ${\bf V}_{g+1}$ 
\beq\label{vg3}
\left\{
{\rm Det}\left[G (1-H) \right]
\right\}^{-1/2}
\exp{\left[\frac 12 (B_2,B_1) (1-H)^{-1} G^{-1}
\left(\begin{array}{c}
C_2\\C_1
\end{array}\right)
\right]}~,
\eeq
where the determinant is over all indices (space-time, loop and oscillator
indices). The definitions of $B$, $C$ and $H$ are: 
\begin{equation} \label{defv}
H = 
\left(\begin{array}{cc}
D_{nm}(U_\mu\tilde{\bar{V}}_\nu) {\cal S}_\nu &
- D_{nm}(U_\mu {V}_\nu) \\
- D_{nm}(\tilde{\bar{U}}_\mu\tilde{\bar{V}}_\nu) 
{\cal S}_\mu^{-1} {\cal S}_\nu &
 D_{nm}(\tilde{\bar{U}}_\mu {V}_\nu) 
{\cal S}_\mu^{-1}
\end{array}\right)
\end{equation}
$$
\begin{array}{l}
(B_1)_m^\nu \!= \!\!\!
\sum\limits_{\begin{subarray}{l} 1\le \mu\le g\\ 1\le k<\infty\end{subarray}}\!\!\!\! 
a_0^\mu G \left[-D_{0k}(U_\mu) +
D_{0k}(\tilde{\bar{U}}_\mu) {\cal S}^{-1}_\mu +
\delta_{\mu\nu} D_{0k}(U_\mu) \right] D_{km}(V_\nu)
+a_0^\nu G D_{0m}(V_\nu)
\nonumber \\ 
(B_2)_m^\nu \!= \!\!\!
\sum\limits_{\begin{subarray}{l} 1\le \mu\le g\\ 1\le
    k<\infty\end{subarray}}\!\!\!\! 
a_0^\mu G \left[D_{0k}(U_\mu) -
D_{0k}(\tilde{\bar{U}}_\mu) {\cal S}^{-1}_\mu +
\delta_{\mu\nu} D_{0k}(\tilde{\bar{U}}_\mu) 
{\cal S}^{-1}_\mu \right] D_{km}(\tilde{\bar{V}}_\nu)
{\cal S}_\nu + a_0^\nu G D_{0m}(\tilde{\bar{V}}_\nu)
\nonumber \\  
(C_2)_m^\nu \!= \!\!\!
\sum\limits_{\begin{subarray}{l} 1\le \mu\le g\\ 1\le
    k<\infty\end{subarray}}\!\!\!\! 
D_{mk}(U_\nu) G
\left[-D_{k0}(V_\mu) + {\cal S}_\mu 
D_{k0}(\tilde{\bar{V}}_\mu)  +
\delta_{\mu\nu} D_{k0}(V_\mu) \right] a_0^\nu
+ D_{m0}(U_\nu) G a_0^\nu  
\nonumber \\ 
(C_1)_m^\nu \!= \!\!\!
\sum\limits_{\begin{subarray}{l} 1\le \mu\le g\\ 1\le
    k<\infty\end{subarray}}\!\!\!\! 
D_{mk}(\tilde{\bar{U}}_\nu) G 
{\cal S}_\nu^{-1} \left[D_{k0}(V_\mu) -
D_{k0}(\tilde{\bar{V}}_\mu) {\cal S}_\mu +
\delta_{\mu\nu} D_{k0}(\tilde{\bar{U}}_\mu) 
{\cal S}_\mu \right] a_0^\mu 
+ D_{m0}(\tilde{\bar{U}}_\nu) G a_0^\nu  .
\nonumber
\end{array}
$$ 
Notice that in the expressions for $B_2$ and $C_1$, that follow
from~\eq{vg2} and~\eq{D-rep}, using also ~\eq{df-cl} , the string
propagators $P$ contained in the tilded local coordinates are not
always paired with the corresponding space-time matrix ${\cal
S}$. This is the origin of the subtlety mentioned in
Section~\eq{clo-chan}. At this point one needs to combine the
result~\eq{vg3} with the zero-mode part of Eq.~\eq{vg1} and keep
following the steps of Appendix~D of~\cite{DiVecchia:1988cy}, now also
with the space-time matrices ${\cal S}$. In the present case the
combination of local coordinate representing the generators of the
Schottky group is $S_\mu^{\rm cl} \equiv \tilde{\bar{V}}_\mu
U_\mu$. By using many times the group property~\eq{D-rep} we arrive to
formulae analogous to~(D.21) and then~(D.24)
of~\cite{DiVecchia:1988cy}. In particular, we find that the matrix
$C_{\mu\nu}^{(1)}$ in our~\eq{zm-cl} is
\beqa
C^{(1)}_{\mu\nu}&=&\left(D_{00}(S_\mu)+D_{00}(S^{-1}_\mu\right))\delta_{\mu\nu}
\label{D24}\\
&+&
{}^{(-\mu)}{\sum_{\alpha}}^{{}_{\scriptstyle (-\nu)}}
D_{0n}(S_\mu)  D_{nm}(T_\alpha) 
D_{m0}({S}_\nu) {\cal T}_{\a} {\cal S}_{\nu}
\nonumber \\ &-& 
{}^{(-\mu)}{\sum_{\alpha}}^{\prime{}_{\scriptstyle (+\nu)}}
D_{0n}(S_\mu)  D_{nm}(T_\alpha) 
D_{m0}({S}^{-1}_\nu) {\cal T}_{\a}
\nonumber \\
&-&{}^{(+\mu)}{\sum_{\alpha}}^{\prime {}_{\scriptstyle (-\nu)}}
D_{0n}(S^{-1}_\mu)  D_{nm}(T_\alpha) 
D_{m0}({S}_\nu) {\cal S}_{\mu}^{-1} {\cal T}_{\a} {\cal S}_{\nu}
\nonumber \\ &+&  
{}^{(+\mu)}{\sum_{\alpha}}^{{}_{\scriptstyle (+\nu)}}
D_{0n}(S^{-1}_\mu)  D_{nm}(T_\alpha) 
D_{m0}({S}^{-1}_\nu) {\cal S}_{\mu}^{-1} {\cal T}_{\a} ~,
\nonumber
\eeqa
where now the indices $n,m$ run from $1$ to~$\infty$ and the superscript
``cl'' on the elements $S_\mu$ and $T_\a$ of the Schottky groups is
understood. The sums with the superscripts ${(\pm\mu)}$ at the left and
${(\pm\nu)}$ at the right are taken over all elements $T_{\alpha}$ of the
Schottky group which contain neither $S^{\pm n}_\mu$ as leftmost factor nor
$S^{\pm n}_\nu$ as rightmost factor for any {\em positive} integer $n$;
moreover the prime on the sum means that the identity is absent when
$\mu=\nu$.  Finally the matrix ${\cal T}_{\a}$ is the space-time analogue of
the Schottky element $T_\a$ and is the product of the ${\cal S}_{\mu}$
corresponding to the generators contained in $T_\a$.  Then by using the
explicit expressions for the $D$'s~\eq{D}, one can check that~\eq{D24} is just
the series expansion of the following logarithms
\beqa\label{C15}
C^{(1)}_{\mu\nu}&=& -\delta_{\mu\nu} \ln (a_\mu d_\mu) +
{}^{(-\mu)}{\sum_{\alpha}}^{{}_{\scriptstyle (-\nu)}}
\ln\left[\frac{S_\mu^{-1}(\infty)-T_\a(0)}{S_\mu^{-1}(\infty)-T_\a S_\nu(0)}
 \right] {\cal T}_{\a} {\cal S}_{\nu}
\nonumber \\ & - & 
{}^{(-\mu)}{\sum_{\alpha}}^{\prime{}_{\scriptstyle (+\nu)}}
\ln\left[\frac{S_\mu^{-1}(\infty)-T_\a(0)}{S_\mu^{-1}(\infty)-T_\a
    S_\nu^{-1}(0)}  \right] {\cal T}_{\a} 
\nonumber \\ & - & 
{}^{(+\mu)}{\sum_{\alpha}}^{\prime {}_{\scriptstyle (-\nu)}}
\ln\left[\frac{S_\mu(\infty)-T_\a(0)}{S_\mu(\infty)-T_\a S_\nu(0)}
\right] {\cal S}_{\mu}^{-1} {\cal T}_{\a} {\cal S}_{\nu}
\nonumber \\ & + &  
{}^{(+\mu)}{\sum_{\alpha}}^{{}_{\scriptstyle (+\nu)}}
\ln\left[\frac{S_\mu(\infty)-T_\a(0)}{S_\mu(\infty)-T_\a S_\nu^{-1}(0)}
\right] {\cal S}_{\mu}^{-1} {\cal T}_{\a}~.
\eeqa
Finally one can check that this expression agrees with the one
of~\eq{zm-cl} given in terms of integrals of the differentials
$\bm{\zeta}_\nu$'s when the base point is placed at infinity
$w=\infty$. However, as proved in the main text, that combination of
integrals is independent of this particular choice, so finally we can
write $C^{(1)}_{\mu\nu}$ as in Eq.~\eq{zm-cl}.

\end{document}